\documentclass[prb,showpacs,superscriptaddress,twocolumn,10pt]{revtex4-1}

\usepackage{mathrsfs}
\usepackage{times}
\usepackage{tikz}
\usepackage{amsfonts}		
\usepackage{amsmath}
\usepackage{amssymb}
\usepackage{wasysym}
\usepackage{marvosym}
\usetikzlibrary{arrows,shapes}

\usepackage{color}
\usepackage{graphicx}
\usepackage{epsfig}				
\usepackage{subfigure}
\usepackage[utf8]{inputenc}
\usepackage{lmodern}
\usepackage[normalem]{ulem}
\usepackage[colorlinks]{hyperref}

\hypersetup{colorlinks, citecolor=blue, linkcolor=red, urlcolor=blue}

\newcommand{\orcid}[1]{\href{https://orcid.org/#1}{\includegraphics[width=7pt]{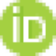}}}

\begin{document}

\title{NMR Relaxation by Redfield's equation in a spin system $I=7/2$}

\author{A. Consuelo-Leal\orcid{0000-0003-1141-210X}}\email{adrianeleal25@gmail.com}
\affiliation{Instituto de F\'{i}sica de S\~{a}o Carlos, Universidade de S\~{a}o Paulo, CP 369, 13560-970, S\~{a}o Carlos, S\~{a}o Paulo, Brasil.}

\author{A. G. Araujo-Ferreira\orcid{0000-0002-6676-384X}}
\affiliation{Instituto de F\'{i}sica de S\~{a}o Carlos, Universidade de S\~{a}o Paulo, CP 369, 13560-970, S\~{a}o Carlos, S\~{a}o Paulo, Brasil.}

\author{E. Lucas-Oliveira\orcid{0000-0003-1353-918X}}
\affiliation{Instituto de F\'{i}sica de S\~{a}o Carlos, Universidade de S\~{a}o Paulo, CP 369, 13560-970, S\~{a}o Carlos, S\~{a}o Paulo, Brasil.}

\author{T. J. Bonagamba\orcid{0000-0001-8894-9170}}
\affiliation{Instituto de F\'{i}sica de S\~{a}o Carlos, Universidade de S\~{a}o Paulo, CP 369, 13560-970, S\~{a}o Carlos, S\~{a}o Paulo, Brasil.}

\author{R. Auccaise\orcid{0000-0002-9602-6533}}\email{raestrada@uepg.br}
\affiliation{Departamento de F\'{i}sica, Universidade Estadual de Ponta Grossa, Av. General Carlos Cavalcanti 4748, 84030-900 Ponta Grossa, Paran\'{a}, Brasil.}

\date{\today}

\begin{abstract}
Redfield's master equation is solved analytically for a nuclear system with spin $I=7/2$. The solutions of each density matrix element are computed using the irreducible tensor operator basis.  The $^{133}$Cs nuclei of the caesium-pentadecafluorooctanoate molecule in a lyotropic liquid crystal sample at the nematic phase and at room temperature was used as an experimental setup. Experimental longitudinal and transverse magnetization dynamics of the $^{133}$Cs nuclei signal were monitored and by numerical procedures the theoretical approach generates valuable mathematical expressions with the highest accuracy. The methodology introduced could be extended without major difficulties to other nuclei species.
\end{abstract}

\pacs{Valid PACS appear here}
\maketitle

\section{Introduction}

Since it was established that relaxation processes are the basis of some medical imaging technologies, relaxation processes on nuclear spin systems have extended to other ranges of research:  gas-oil and food. Mainly, the relaxation technique exploits hydrogen nuclei characteristics, spin value $I=1/2$, $100\%$ natural abundance, and the highest gyromagnetic ratio. On the other hand, recent advances exploring  $^{209}$Bi  nuclei with  $I=9/2$ and $100\%$ natural abundance have been highlighted as promising organometallic compounds for magnetic resonance imaging \cite{gosweiner2018,kruk2019JCP}. Similarly,  $^{14}$N  nuclei with $I=1$ and very low natural abundance were exploited by the quadrupole relaxation enhancement on solid proteins monitoring the $^{1}$H-$^{14}$N relaxation dynamics \cite{kruk2019B}; also, performing the fast field cycling NMR relaxometry to identify molecular dynamics  influenced by $^{14}$N chemical environments \cite{conte2021} and the  food industry in general \cite{ates2021}. The common nuclear characteristic between those recent advances is the spin value $I>1/2$; which inevitably entails the emergence of the nuclear quadrupole moment definition and its introduction to relaxation processes.

Nuclear quadrupole relaxation phenomena have been investigated through various approaches to our knowledge: transition probability theory \cite{van-kranendonk1954},  transition probability using perturbation theory \cite{yosida1956,kochelaev1960,andrew1961},   Heitler-London approximation \cite{kondo1959},  tight-binding approximation \cite{narath1967}, and  Redfield's theory \cite{jaccard1986} all those performed at many different materials, spin values,  and nuclear species. The  dynamics of longitudinal and transverse magnetization for spin values $I>1/2$ were modeled by a multiexponential decay description, which is a consensual point of view \cite{jaccard1986,kelly1992,mcdowell1995}.  Since those methods were established, the most commonly applied was the transition probability using perturbation theory  \cite{gordon1978A, suter1988, mcdowell1995, sarthour2003, lim2012, yesinowski2015}, which has the most straightforward access to information from the experimental data.

The other one,  Redfield's theory, has been explored to establish differential equations of some operator mean values that allow extracting analogous information from the previous methods and, in some cases, without solving the differential equation \cite{mcconnell1987Book}. The procedure's main purpose is to identify the relaxation rates that govern the nuclear spin dynamics, which are identified as coefficients that multiply each operator's mean value factor \cite{mcconnell1987Book,abragam1994Book,kruk2016Book,kowalewski2018Book}. 

A parallel approach, still applying Redfield's theory, has been performed to compute differential equations for each density matrix element  and establish $2I+1$ linear systems of differential equations, each one for each coherence order \cite{jaccard1986}. An advantage of this approach was highlighted during the measurements of interatomic distances between nuclei with quadrupole moments when higher coherence orders were computed \cite{eliav1998}, also quantifying correlation times when some sets of coherence orders were monitored under the relaxation dynamics \cite{auccaise2008}.

Specifically, experimental procedures exploring relaxation processes were reported on $I=7/2$ spin systems. In soft matter, the central transition of $^{139}$La nuclear species relaxation times on protein samples of bovine serum albumin was characterized \cite{reuben1976}. Only the central line was monitored because the frequency distance between the first satellite line and the central line was approximately 7.5 MHz, which is considered a  large quadrupole coupling.  It implies that the transition of energy levels $\frac{+1}{2}\leftrightarrow \frac{-1}{2}$ are selectively excited without perturbing the other ones. In solid state, the central transition of the  $^{59}$Co signal in a single crystal of K$_{3}$Co(CN)$_{6}$ molecule at the temperature of 295 K and quadrupole coupling of 300 kHz was monitored \cite{gordon1978B}.  Furthermore, when different temperature values are considered, i.e., different molecule agitation levels, some optical modes can be identified as a source of relaxation phenomena \cite{hoch1976}. Another study detecting $^{133}$Cs signals of a Cs$_{3}$D(SO$_{4}$)$_{2}$ single crystal at a variable temperature between 180 K and 430 K was performed. The main purpose of the study was identifying phase transitions dependent on temperature \cite{lim2012}.  The common theoretical assumption of these studies is about the partial solution of Redfield's equation \cite{reuben1976,hubbard1970B}, meaning that only the antisymmetric functions of the basis set will contribute to the standard observables of magnetization \cite{reuben1976}.

On the other hand, from the fundamentals of quantum information processing, an interesting approach arises to explore the complete solution of Redfield's equation. These solutions will define the quantum dynamics used to come back through the steady state. Therefore, in this study, we develop a theoretical analysis and introduce a methodology to explore the advantages of the complete solutions of the master equation for a large spin system. To contextualize the relaxation dynamics  of a nuclear system of large spin value, we use $I=7/2$. This paper is organized to explore the complete solution of Redfield's equation as follows. First, in Sec. \ref{sec:RedfieldTheoryForANuclearSystem} is established the main ideas on Redfield's master equation following the criteria of A. Abragam \cite{abragam1994Book} and J. McConnell \cite{mcconnell1987Book}, emphasizing the irreducible tensor operator basis. Next, in Sec. \ref{sec:DensityMatrixElementsSolutions}, the density matrix notation is introduced and the solutions are  established. Appendix \ref{app:SolvingRedfieldEquation} contains an extended explanation. Then, in Sec.  \ref{sec:TheoreticalApplicationOn133Cs} and Sec. \ref{sec:ExperimentalDescriptionOn133Cs}, theoretical and experimental applications are introduced and detailed, respectively.   Finally, the most important remarks on the data analysis are discussed in Sec. \ref{sec:Discussions}, and the conclusions are summarized in \ref{sec:Conclusions}. 

\section{Redfield's theory for a nuclear system}
\label{sec:RedfieldTheoryForANuclearSystem} 

The dynamics of an open quantum system such as any nuclear spin system are formally described by  Redfield's theory \cite{jaccard1986, levitt2001Book, Segnorile2006, kruk2016Book}, and extended to other approaches \cite{Sykes2012,Thingna2013,Purkayastha2017,Lindner2018,FarinaDonato2019}. A standard introduction has been made by using the second-order approximation of the Liouville-von Neumann equation \cite{redfield1957}, generating an equation of motion in terms of the double commutator operator on one hand and a time derivative operation on the other. Some applications devoted to analyzing the time evolution of density matrix elements, provide a concise notation as described in Eq. (3.58) of Ref. \cite{mcconnell1987Book}. It is reproduced in Eq. (\ref{EquacaoRedfield}), considering the time evolution of each element  $\rho_{\alpha, \beta}$  as denoted by  the following differential equation: 
\begin{equation}
\frac{\partial {\rho }_{\alpha, \beta}\left( t\right) }{\partial t%
}=\sum\limits_{ \alpha ^{\prime } \beta ^{\prime }}\mathcal{R}_{\alpha \beta}^{\alpha ^{\prime }
\beta ^{\prime }}e^{i\left( \omega _{\alpha }-\omega _{ \alpha^{\prime } }+\omega
_{\beta ^{\prime }}-\omega _{\beta}\right) t}{\rho }_{\alpha ^{\prime },
\beta ^{\prime }}\left( t\right),  \label{EquacaoRedfield}
\end{equation}
where  $\mathcal{R}_{\alpha\beta}^{ \alpha ^{\prime } \beta
^{\prime }}$ represents the  transformation  element $\alpha \beta \alpha ^{\prime }\beta
^{\prime }$ of the relaxation superoperator with $\alpha ,\alpha ^{\prime
},\beta ,\beta ^{\prime }$, and $\omega_{\alpha}$, $\omega_{\alpha
^{\prime}}$, $\omega_{\beta }$, $\omega_{\beta ^{\prime}}$ represent the angular frequencies corresponding to  the Zeeman  eigenenergies \cite{abragam1994Book,redfield1957}.
The relaxation superoperator depends on the specific interactions between the spin system and a thermal reservoir which are characterized by spectral density functions and correlation times of the local field fluctuations. Particularly, in the present analysis, the quadrupolar interaction must be considered as the main relaxation mechanism \cite{mcconnell1987Book,abragam1994Book}.

The most standard procedure to progress with Redfield's theory is applying the  mean value  of the operator ${\mathbf{I}}_{r}$  associated with the Cartesian elements of magnetization, i.e. $\left\langle {\mathbf{I}}_{r}\right\rangle$ with $r=x,y,z$. The theoretical development uses  the properties of commutation rules between angular momentum operators, the linear algebra definitions applied to analyze the matrix representation of operators, and the fundamentals of linear systems of differential equations. From the logical execution of those procedures, an appropriate differential equation was established as defined in Eq. (9.25) of Ref. \cite{mcconnell1987Book},  and 2nd-rank operators as denoted by Eq. (9.23) of Ref. \cite{mcconnell1987Book}, such that for spin $I=1$ the dynamics of the transverse and longitudinal magnetization were successfully computed \cite{mcconnell1987Book,abragam1994Book}. On those applications, the Redfield's equation at its commutator notation was rewritten by 
\begin{equation}
\frac{1}{C}\frac{d\left\langle {\mathbf{I}}_{r}\right\rangle }{dt} = - \sum_{p=-2}^{2}  \left( -1 \right)^{p} J_{p}\left\langle \left[ {\mathbf{Q}}_{p},\left[ 
{\mathbf{Q}}_{-p},{\mathbf{I}}_{r}\right] \right] 
\right\rangle\text{,}   \label{RelaxationDensityMatrixElementsCorrigida}
\end{equation}
where $J_{p}\equiv J \left( p \omega_{0} \right)$ with $p=0,1,2$ represent the spectral density functions, 
the operators  ${\mathbf{Q}}_{p}$ with $p=0, \pm 1, \pm 2$ denote second rank irreducible tensor operators \cite{jacobsen1976,mcconnell1987Book} for any spin value $I$. The proportionality constant parameter $C$ is denoted by 
\begin{equation}
C=\frac{9}{10}\frac{1}{\left( 2I\left( 2I-1\right) \right) ^{2}}\left( \frac{ eQ}{\hbar }\mathcal{V}_{z z }  \right) ^{2}\left( 1+\frac{\eta ^{2}}{3}\right) \text{,}
\label{ConstantProportionalityGeneral}
\end{equation}
for any spin value $I$, $e$ means the elemental charge, $Q$ means the quadrupole moment of the nucleus, $\mathcal{V}_{z z }$ means the electric field gradient along the $z$-axis, $\eta$ means the asymmetry parameter with $1\geq \eta \geq 0$, and $ \hbar $ means the reduced Planck's constant. The parameter $C$ satisfies the Eq. (5) and Eq. (72) of Ref.  \cite{jaccard1986} for spin $I=3/2$ and  spin $I=5/2$, respectively.

An alternative procedure to explore the full advantage of Redfield's theory consists in the computation of the density matrix elements in terms of irreducible tensor operators \cite{jaccard1986}, with  $ {\mathbf{I}}_{r} = {\mathbf{T}}_{l,m} $. In this sense, the mathematical procedure is developed in some stages. The first stage consists in applying specific algebraic procedures on Eq. (\ref{RelaxationDensityMatrixElementsCorrigida}), and the generation of a set of linear systems of differential equations using the order of coherence criteria, i.e. $l$th coherence order with  $l=0,1,\ldots,2I$. The second stage consists in finding solutions of a linear system using properties of matrices and procedures to solve time-dependent first order linear differential equations of one variable. The main characteristic of this second stage is that the dimension of the linear system satisfies the relation  $2I+1 - q $, where $q$ is the coherence order. Therefore, the set of linear systems to solve are the coherences of intermediate and higher coherence order values. The execution of both stages allows computing the $p$-characteristic relaxation rates of each $q$th-coherence order, $R _{p}^{\left( q\right) } $. The most challenging task is the linear system solution for the lower coherence orders, which implies on a greater linear system dimension and finding an analytical solution sometimes is a hard problem.

Particularly, for the case of $I=7/2$, a standard notation of these operators ${\mathbf{Q}}_{r}$  is writing them in terms of  spin angular momentum operators (see Eq. (9.22) and Eq. (9.23) of Ref. \cite{mcconnell1987Book}, or Eq. (23) on page 233 of Ref. \cite{abragam1994Book}), largely used for theoretical developments on spectroscopy. Another convenient representation for the  ${\mathbf{Q}}_{r}$ operators is  using the irreducible tensor operator basis, given by $ {\mathbf{I}} _{x} = \frac{\sqrt{84}}{2}\left({\mathbf{T}}_{1,-1} - {\mathbf{T}} _{1,+1}\right)$ and  $ {\mathbf{I}} _{z} = \sqrt{42} {\mathbf{T}}_{1,0}$. Consequently, the ${\mathbf{Q}}_{r}$ operators are denoted by
\begin{eqnarray}
{\mathbf{Q}}_{\mp 2} &=&42\sqrt{6}\left({\mathbf{T}}_{1,\pm 1}\right) ^{2}%
\text{,}  \label{QuadrupolarBathSecondMinus7o2} \\
{\mathbf{Q}}_{\mp 1} &=&-42\sqrt{3}\left( {\mathbf{T}}_{1,0}{\mathbf{%
T}}_{1,\pm 1}+{\mathbf{T}}_{1,\pm 1}{\mathbf{T}}_{1,0}\right) \text{,}
\label{QuadrupolarBathFirstMinus7o2} \\
{\mathbf{Q}}_{0} &=&63\left( 2 \left({\mathbf{T}}_{1,0}\right) ^{2}-%
\frac{\sqrt{7}}{4}{\mathbf{T}}_{0,0}\right) \text{.} \label{QuadrupolarBathZero7o2} 
\end{eqnarray}

The main advantage of the use of the irreducible tensor operator basis is that it directly relates to the  coherence orders of the density matrix elements. That implies that the solutions of Redfield's equations are solved between elements of the same order, and there is no mixture between elements of different orders.

\section{Density matrix elements solutions}
\label{sec:DensityMatrixElementsSolutions}

The relaxation dynamics of a nuclear spin system coupled weakly with an environment is well established, finding the solution to  Redfield's master equation \cite{abragam1994Book}. In particular, some  theoretical analysis have been made in nuclear spins systems  with $I=1,3/2,5/2$ detailed in Ref. \cite{jaccard1986,kelly1992,mcconnell1987Book}, for spin $I=2$ detailed in Ref.  \cite{korblein1985} and also in $I=7/2$ in Ref. \cite{gordon1978A,tsoref1996}, such that, these references are used as guidelines for the current study. Let us denote  the density operator,  for any spin value $I$,  using the bracket notation
\begin{equation}
{\boldsymbol{\rho }}=\sum_{\alpha=1}^{2I+1}\sum_{\beta=1}^{2I+1}\rho _{\alpha,\beta}\left\vert
I+1-\alpha\right\rangle \left\langle I+1-\beta\right\vert \text{.}
\end{equation}
In such wise, the particular case of the current application for spin value with $I=7/2$,  its matrix  representation has frequently been denoted by
\begin{equation}
{\boldsymbol{\rho }} = \left[ 
\begin{array}{cccccccc}
\rho _{1,1} & \rho _{1,2} & \rho _{1,3} & \rho _{1,4} & \rho _{1,5} & \rho
_{1,6} & \rho _{1,7} & \rho _{1,8} \\ 
\rho _{2,1} & \rho _{2,2} & \rho _{2,3} & \rho _{2,4} & \rho _{2,5} & \rho
_{2,6} & \rho _{2,7} & \rho _{2,8} \\ 
\rho _{3,1} & \rho _{3,2} & \rho _{3,3} & \rho _{3,4} & \rho _{3,5} & \rho
_{3,6} & \rho _{3,7} & \rho _{3,8} \\ 
\rho _{4,1} & \rho _{4,2} & \rho _{4,3} & \rho _{4,4} & \rho _{4,5} & \rho
_{4,6} & \rho _{4,7} & \rho _{4,8} \\ 
\rho _{5,1} & \rho _{5,2} & \rho _{5,3} & \rho _{5,4} & \rho _{5,5} & \rho
_{5,6} & \rho _{5,7} & \rho _{5,8} \\ 
\rho _{6,1} & \rho _{6,2} & \rho _{6,3} & \rho _{6,4} & \rho _{6,5} & \rho
_{6,6} & \rho _{6,7} & \rho _{6,8} \\ 
\rho _{7,1} & \rho _{7,2} & \rho _{7,3} & \rho _{7,4} & \rho _{7,5} & \rho
_{7,6} & \rho _{7,7} & \rho _{7,8} \\ 
\rho _{8,1} & \rho _{8,2} & \rho _{8,3} & \rho _{8,4} & \rho _{8,5} & \rho
_{8,6} & \rho _{8,7} & \rho _{8,8}
\end{array}
\right] \text{.} \label{DensityMatrix8x8}
\end{equation}
In this analysis, a standard mathematical development is performed to compute the solution for the density matrix elements $\rho _{\alpha,\beta}$ with $\alpha \geqslant \beta$, such that the solutions for $\alpha<\beta$ are generated by the hermitian property of the density operator. {However, the analysis developed to solve the problem is a very long and tedious mathematical process. To put in evidence this hard procedure some algebraic routines were implemented and many of them are detailed and extensively described in Appendix \ref{app:SolvingRedfieldEquation}. Probably the most relevant result of the mathematical process is a concise notation of each density matrix element of Eq. (\ref{DensityMatrix8x8}) as denoted by }
\begin{equation}
\rho _{q+n,n}\left( t\right)   =\rho^{\text{eq}}_{q+n,n}+ \sum_{p=1}^{p_{\text{max}}} \overline{W}_{n,p }^{\left( q \right) }\exp \left[ -R_{p}^{\left( q \right) }\left( t-t_{0}\right) \right] \widetilde{\rho }_{p}\left( t_{0}\right)  \text{,}  \label{SolutionDensityMatrixElements}
\end{equation}
where  $p$ and $n$ are dummy indexes bounded by  $p_{\text{max}},n_{\text{max}}=2I+1-q$, $R_{p}^{\left( q \right) } = - C \lambda_{p}^{\left( q\right)} $ mean the $p$th relaxation rate of the $q$th coherence order, $ \overline{W}_{n,p }^{\left( q \right) }$ represents each matrix element of the inverse transformation $\overline{\mathbf{W}}^{ \left( q \right)}$ (the overline symbol means the inverse transformation),  $\widetilde{\rho }_{p}\left( t_{0}\right) $ denotes the initial condition defined by the Eq. (\ref{RedfieldElementoGeralInitialConditions}). Also, in Tab. \ref{tab:Eigenvalues} are displayed the eigenvalues of each Redfield's superoperator $\boldsymbol{\mathcal{J}}^{\left( q\right)}$ computed for 2nd, 3rd,..., 7th coherence order. The case of zero and first coherence order are detailed introducing extended explanations for a better understanding. To put in evidence the applicability of the solutions, we introduce one theoretical application and an experimental one.

\section{Theoretical application}
\label{sec:TheoreticalApplicationOn133Cs}

The density matrix evolution  of any spin system with $I=7/2$ interacting with an environment, it represented by classical fields in this study, can be described by the Redfield's equation. Let us consider the initial quantum state as denoted by $\left\vert \Psi \right\rangle= \left\vert 7/2\right\rangle + \left\vert -7/2\right\rangle$, avoiding the normalization constants, so that the density operator is denoted by
\begin{equation}
{\boldsymbol{\rho }} \left( t_{0}\right) = \frac{\left( \left\vert 7/2 \right\rangle
+\left\vert -7/2\right\rangle \right)   \left(
\left\langle 7/2\right\vert +\left\langle %
-7/2\right\vert \right)}{2} \text{,}
\end{equation}
or using the  matrix element notation,  $\rho _{1,1}\left( t_{0}\right)$, $\rho _{1,8}\left( t_{0}\right)$, $\rho _{8,1}\left( t_{0}\right)$, and $\rho _{8,8}\left( t_{0}\right)$ should assume non-null values. The formal model along the time evolution of these elements can be computed using Eq. (\ref{SolutionDensityMatrixElements}) for $q=0$ and $q=7$. Therefore, these four elements are described by
\begin{eqnarray}
\rho _{1,1}\left( t\right)   &=& \rho^{\text{eq}}_{1,1}+ \sum_{p=1}^{8}\overline{W}_{1,p}^{ \left( 0 \right) }\exp \left[ -R_{p}^{\left( 0 \right) }\left( t\right) \right] \widetilde{\rho }_{p}\left( {0}\right)  \text{,} \label{rho11NooNState}\\
\rho _{8,8}\left( t\right)   &=&\rho^{\text{eq}}_{8,8}+ \sum_{p=1}^{8} \overline{W}_{8,p}^{ \left( 0 \right)  }\exp \left[ -R_{p}^{\left( 0 \right) }\left( t\right) \right] \widetilde{\rho }_{p}\left( {0}\right)  \text{,}\label{rho88NooNState}  \\
\rho _{8,1}\left( t\right)   &=& \sum_{p=1}^{1} \overline{W}_{1,p}^{ \left( 7 \right)  }\exp \left[ -R_{1}^{\left( 7 \right) }\left( t\right) \right] \widetilde{\rho }_{p}\left( {0}\right)  \text{,}\label{rho81NooNState} \\
\rho _{1,8}\left( t\right)   &=& \rho _{8,1}^{*}\left( t\right)  \text{.}\label{rho18NooNState}
\end{eqnarray}
Without loss of generality we compute the spectral density values assuming the isotropic motion \cite{jaccard1986,abragam1994Book} at the Larmor frequency $\omega_{L}/2\pi  = 47.24$ MHz, quadrupolar frequency  $\omega_{Q}/2\pi  = 266$ kHz  and  the correlation time $\tau_{c} = 4.1 $ ns reported on Ref. \cite{tsoref1996}, such that  $J_{0}=8.2\times 10^{-9}$s, $J_{1}=3.3\times 10^{-9}$s, $J_{2}=1.2\times 10^{-9}$s, and $C=2.793\times 10^{11}$ Hz$^{2}$. Another versatility of the  solutions highlight the choice of the temperature regime of the steady state, here defined at a low temperature regime denoted by 
\begin{equation}
{\boldsymbol{\rho }} ^{\text{eq}} =   \left\vert 7/2\right\rangle    \left\langle 7/2\right\vert \text{.}
\end{equation}

\begin{table}[t]
\begin{center}
\begin{ruledtabular}
\begin{tabular}{c c r r r r}
$q$	& $k$ &  $\overline{W}_{1,k}^{ \left( q \right) }\widetilde{\rho }_{k}\left( {0}\right)$  &  $\overline{W}_{8,k}^{ \left( q \right) }\widetilde{\rho }_{k}\left( {0}\right)$  &  $R_{k}^{\left( q\right)} $	(kHz)  & T$_{k}^{\left( q\right)}$	($\mu$s) \\     \cline{1-6}   
& 1  & $-0.07$	&  $0.07$	&  $56.98$ & $17.55$ \\ 
& 2  & $-6.27$	&  $6.27$	&   $4.13$ & $242.37$ \\ 
& 3  & $-0.60$	&  $0.60$	&  $28.32$ & $35.31$ \\ 
0& 4 & $-0.06$	&  $0.06$	&  $15.85 $&$ 63.09$ \\ 
& 5  &	$0$		&	$0$		&  $56.41$ & $17.73$ \\ 
& 6  &	$0$		&	$0$		&   $34.04$ & $29.37$ \\ 
& 7  &	$0$		&	$0$		&  $ 14.82 $& $67.48$ \\ 
& 8  &	$0$		&	$0$		&    $0$     & $\infty$		\\ \hline
7& 1 &	$0.50$		&			&  $ 21.69$ & $46.10$ \\
\end{tabular} 
\end{ruledtabular}
\end{center} \caption{Zero and seventh coherences order amplitudes $\overline{W}_{k^{\prime},k}^{ \left( q \right) }\widetilde{\rho }_{k}\left( {0}\right)$,  relaxation rates $R_{k}^{\left( q\right)} $ and relaxation times T$_{k}^{\left( q\right)}=1/R_{k}^{\left( q\right)}$ are displayed. Eigenvalues of  Redfield's superoperator $\boldsymbol{\mathcal{J}}^{\left( 0\right)}$ and $\boldsymbol{\mathcal{J}}^{\left( 7\right)}$ were used to compute the relaxation rates. } \label{tab:DensityMatrixElementsDynamics}
\end{table}

\begin{figure}[t]
\includegraphics[width=3.40in]{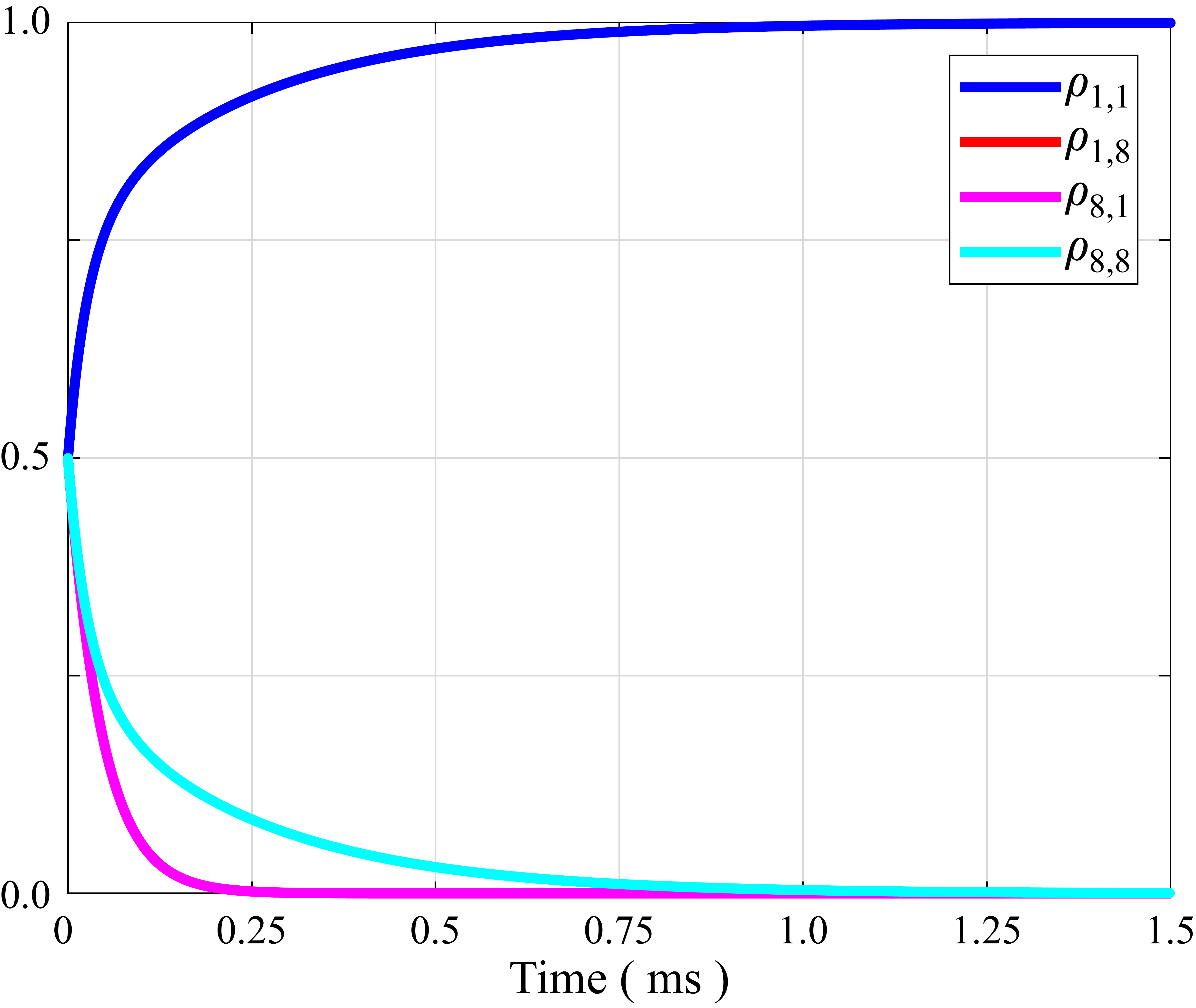} 
\caption[Dynamics]{(Color online) The relaxation dynamics of four density matrix elements defined in Eq. (\ref{rho11NooNState}-\ref{rho18NooNState}) are displayed. Relaxation rates shown in Tab. \ref{tab:DensityMatrixElementsDynamics} are used to monitor the evolution. The evolution of $\rho_{8,1}$, represented by the magenta solid line, overlaps the dynamics of $\rho_{1,8}$, represented by the red solid line.} \label{fig:DensityMatrixElementsDynamics}
\end{figure}

Therefore, eigenvalues were computed from the Redfield's superoperator $\boldsymbol{\mathcal{J}}^{\left( 0\right)}$ and $\boldsymbol{\mathcal{J}}^{\left( 7\right)}$ as detailed in Appendix \ref{app:SolvingRedfieldEquation} and those results are summarized in Tab. \ref{tab:DensityMatrixElementsDynamics}. Furthermore, the coefficients $\widetilde{\rho }_{p}\left( t_{0}\right)$ were computed using Eq. (\ref{RedfieldElementoGeralInitialConditions}) and applied to predict the dynamics of the density matrix elements as displayed in Fig. \ref{fig:DensityMatrixElementsDynamics}. The non-null elements of the initial quantum state have the same value $\rho _{\alpha,\beta}\left( t_{0}\right)=0.5$ as  can be verified in Fig. \ref{fig:DensityMatrixElementsDynamics}, all of them start from the same intensity value. The evolution of  $\rho _{1,1}\left( t\right)$ evolves to the maximum value of the steady state, and the other ones evolve to the null value, as happens with the other density matrix elements not shown. Moreover, considering the intensity data, third and fourth column of Tab. \ref{tab:DensityMatrixElementsDynamics}, the second component of the zero coherence is more dominant than others. Also, comparing the relaxation rates, $R_{2}^{\left(0 \right)} < R_{1}^{\left( 7\right)} $, we assume that the relaxation rate of 7th coherence order element relaxes in a higher rate than the zero coherence order. Those remarks match with the density matrix elements evolution displayed in Fig. \ref{fig:DensityMatrixElementsDynamics}.

\section{Experimental application}
\label{sec:ExperimentalDescriptionOn133Cs}

\begin{figure}[!ht]
\includegraphics[width=3.40in]{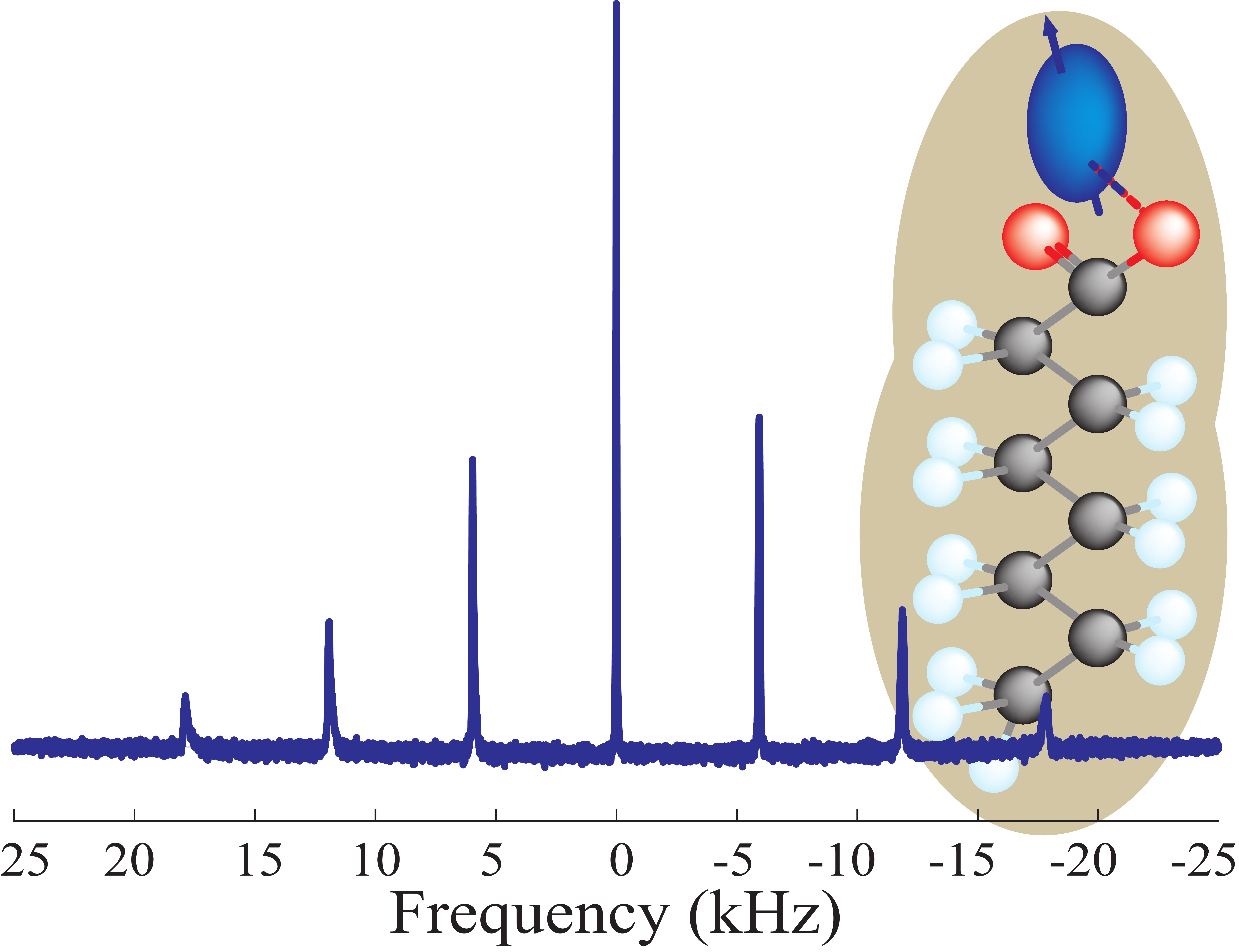} 
\caption[Spectrum and Molecule]{(Color online) The spectrum of $^{133}$Cs nuclei and the cartoon picture of the caesium-pentadecafluooroctanoate (Cs-PFO) are displayed. The blue ellipsoid means  $^{133}$Cs nuclei, red spheres mean  $^{16}$O nuclei, black spheres mean  $^{13}$C nuclei, and blue light spheres mean  $^{19}$F nuclei.} \label{fig:SpectraSamplePFOCs}
\end{figure}

The experimental setup on monitoring the longitudinal and transverse magnetization dynamics was achieved using a Tecmag Discovery Console, a Jastec 9.4 T superconductor magnet and a Jakobsen 5 mm solid-state NMR probe.  A lyotropic liquid crystal sample was placed into a 4 mm o.d. zirconia (ZrO$_{2}$) rotor and sealed with a Kel-F cap. One of the most interesting properties of these lyotropic liquid crystals is the collective orientation capability to achieve appropriate arrangements. From those arrangements, the nematic phase is highlighted by its typical quality of a long-range orientational order of elongated molecules pointing on average in the same direction established by the director   $\textbf{n}$. In this sense, the  caesium-pentadecafluoroctanoate (Cs-PFO) sample {was} used in this experimental development, which is classified at the nematic phase by its stoichiometric composition of Cs-PFO ($40$\%) and   deuterium oxide ($60$\%)   and physical properties \cite{boden1993}. A pictorial scheme about the atomic composition for the Cs-PFO molecule is depicted in Fig.  \ref{fig:SpectraSamplePFOCs}  with molecular formula Cs-C$_{8}$F$_{15}$O$_{2}$.

The  $^{133}$Cs has spin value $I=7/2$ ($100\%$ abundant in nature) ,  such that a finite dimensional Hilbert space is established    which allows establishing the dimension of the Hilbert space  $d=2I+1=8$.   The spectrometer operates at the Larmor frequency of  $\omega_{L}/2\pi =   52.436$ MHz, {and the spectral width used to display each spectrum is $50000$ Hz}. The quadrupolar coupling {of  $\omega_{Q}/2\pi =   5970 \pm 40$ Hz was measured using the frequency distance between }  satellite lines  as can be observed in Fig. \ref{fig:SpectraSamplePFOCs}. {During all experimental implementation,} the sample temperature was fixed at $27.5^{\circ}$C. The $\frac{\pi}{2}$-pulse {length was}   $4.8$ $\mu$s and  the ${\pi}$-pulse length is  $9.4$ $\mu$s. Acquisition time was $81.92$ ms, the number of data points was $4096$, and the dwell time was $20$ $\mu$s. The recycle delay was $15$ s.  The longitudinal and  transverse relaxation times were measured $T_{1}\approx 293  $ ms and  $T_{2}\approx 9.8  $ ms, respectively.

An inversion-recovery pulse sequence \cite{abragam1994Book} 
  was used to monitor the longitudinal magnetization dynamics at $24$ time delays. The experimental data (blue dot symbols) are displayed  in Fig. \ref{fig:MagnetizationZExpThe} using  a vertical scale between -1 and 1.

A Bloch-Echo pulse sequence \cite{abragam1994Book} 
 was used to monitor the transverse magnetization dynamics at $265$ time delays such that each time delay was synchronized at multiple value of the quadrupolar frequency inverse value ($1/\nu_{Q}$). The experimental data (blue dot symbols) are displayed  in Fig. \ref{fig:MagnetizationXExpThe} {using} a vertical scale between 0 and 1.

\section{Discussions}
\label{sec:Discussions} 

The analysis of the data starts establishing the theoretical equations for the longitudinal and transverse magnetization generated by performing Redfield's theory. In this {context},  the definition of the longitudinal magnetization satisfies $\left\langle {\mathbf{I}}_{z}\right\rangle = \mathtt{Tr} \left\{ {\mathbf{I}}_{z} {\boldsymbol{\rho }}  \right\} $ {and}  considering that the experimental data {were}  normalized, the definition must be rescaled by the parameter  $\mathcal{A}_{1}^{z}$. Therefore, the mean value of the $z$ angular momentum operator is denoted by
\begin{eqnarray*}
\left\langle {\mathbf{I}}_{z}\right\rangle & = & \mathcal{A}_{1}^{z} \left( \frac{7}{2} \varrho _{1,1} \left( t \right) + \frac{5}{2} \varrho _{2,2} \left( t \right) + \frac{3}{2}\varrho _{3,3}\left( t\right) +\frac{1}{2}\varrho _{4,4}\left( t \right) \right.  \\
 & & \left. -\frac{1}{2}\varrho _{5,5}\left( t\right) -\frac{3}{2}\varrho_{6,6} \left( t\right) -\frac{5}{2}\varrho _{7,7}\left( t\right) -\frac{7}{2} \varrho _{8,8}\left( t\right) \right) \text{,}
\end{eqnarray*}%
where $\varrho _{k,k}\left( t\right) = \rho _{k,k}\left( t\right)- \rho _{k,k}^{\text{eq}}$.    {Let us consider}  each density matrix element $\rho _{k,k}\left( t\right)$, as denoted by the  Eq. (\ref{SolutionDensityMatrixElements}),  and  each  diagonal matrix element of the    $  {\mathbf{I}}_{z}  $  angular momentum operator, as denoted by  ${I}_{z}   \left( k,k\right) $, such that both of them  are used to   introduce the coefficients $A_{n}^{z}$ as
\begin{equation}
A_{n}^{z} = \sum_{p=1}^{p_{\text{max}}} {I}_{z}   \left( p,p\right)  \overline{W}_{n,p}^{  \left(0\right)   }  \widetilde{\rho }_{n}\left( t_{0}\right) \text{.}
\end{equation}
Also, an equilibrium parameter must be defined that depends on the steady quantum state represented by 
\begin{equation}
A_{z}^{\text{eq}} = \sum_{p=1}^{p_{\text{max}}} {I}_{z}   \left( p,p\right) \rho _{p,p}^{\text{eq}} \text{.}
\end{equation}
Accordingly with both coefficients,  the mean value of the $z$-angular momentum operator component is {rewritten by}
\begin{equation}
\left\langle {\mathbf{I}}_{z}\right\rangle = \mathcal{A}_{1}^{z} \left( A_{z}^{\text{eq}}  +  \sum_{n=1}^{n_{\text{max}}} A_{n}^{z} \exp \left[ -R_{n}^{\left( 0\right) }\left( t-t_{0}\right) \right]  \right)  \text{.}  \label{ExponentialSuperpositionMagZ}
\end{equation}
Moreover,  the initial condition used to solve the differential equation system depends on  the quality of the implementation of the initial quantum state.  For the longitudinal magnetization, the theoretical initial density matrix ${\boldsymbol{\rho }}^{\text{theo}}\left( t_{0}\right)$ must be proportional to  $ -{\mathbf{I}}_{z}$. In this case, the efficiency of the first pulse of the inversion-recovery pulse sequence is encoded by the  parameter   $ \mathcal{A}_{2}^{z} $. {It highlights}  a connection between the theoretical initial state and the experimental initial state, $   {\boldsymbol{\rho }}^{\text{exp}}\left( t_{0}\right) =  \mathcal{A}_{2}^{z} {\boldsymbol{\rho }}^{\text{theo}}\left( t_{0}\right)$. 

In the same way, the definition of the transverse magnetization satisfies $\left\langle {\mathbf{I}}_{x}\right\rangle = \mathtt{Tr} \left\{  {\mathbf{I}}_{x} {\boldsymbol{\rho }}  \right\} $, and considering the  normalization processing of the experimental data defines the parameter  $\mathcal{A}_{1}^{x}$. Therefore, the mean value of the $x$ angular momentum operator is denoted by
\begin{eqnarray*}
\left\langle {\mathbf{I}}_{x}\right\rangle & = & \mathcal{A}_{1}^{x} \left( \sqrt{7}\frac{\rho _{1,2}\left( t\right)+\rho _{2,1}\left( t\right)}{2}+\sqrt{12}\frac{\rho _{2,3}\left( t\right)+\rho _{3,2}\left( t \right)}{2} \right.  \\
 & & \left. +\sqrt{15}\frac{\rho _{3,4}\left( t\right)+\rho _{4,3}\left( t\right)}{2} +4\frac{\rho_{4,5}\left( t\right)+\rho _{5,4}\left( t\right)}{2} \right.  \\  
 & & \left. +\sqrt{15}\frac{\rho _{5,6}\left( t \right)+\rho _{6,5}\left( t \right)}{2}+\sqrt{12}\frac{\rho _{6,7}\left( t \right)+\rho _{7,6}\left( t \right)}{2} \right.  \\
 & & \left. +\sqrt{7}\frac{\rho _{7,8}\left( t\right) +\rho _{8,7}\left( t\right) }{2}\right) \text{.}
\end{eqnarray*}
 The hermitian property  of the density operator, $ \rho _{k,k+1}= \rho _{k+1,k}^{*}$,  enables us  to simplify the mean value mathematical expression
\begin{eqnarray*}
\left\langle {\mathbf{I}}_{x}\right\rangle & = & \mathcal{A}_{1}^{x} \left( \sqrt{7} \rho _{1,2}\left( t\right)+\sqrt{12}\rho _{2,3}\left( t\right) +\sqrt{15}\rho _{3,4}\left( t\right)  \right.  \\
 & & \left. +4\rho_{4,5}\left( t\right) +\sqrt{15}\rho _{5,6}\left( t \right) + \sqrt{12}\rho _{6,7}\left( t \right) \right.  \\
 & & \left. +\sqrt{7}\rho _{7,8}\left( t\right) \right) \text{.}
\end{eqnarray*}

\begin{figure}[t]
\includegraphics[width=3.40in]{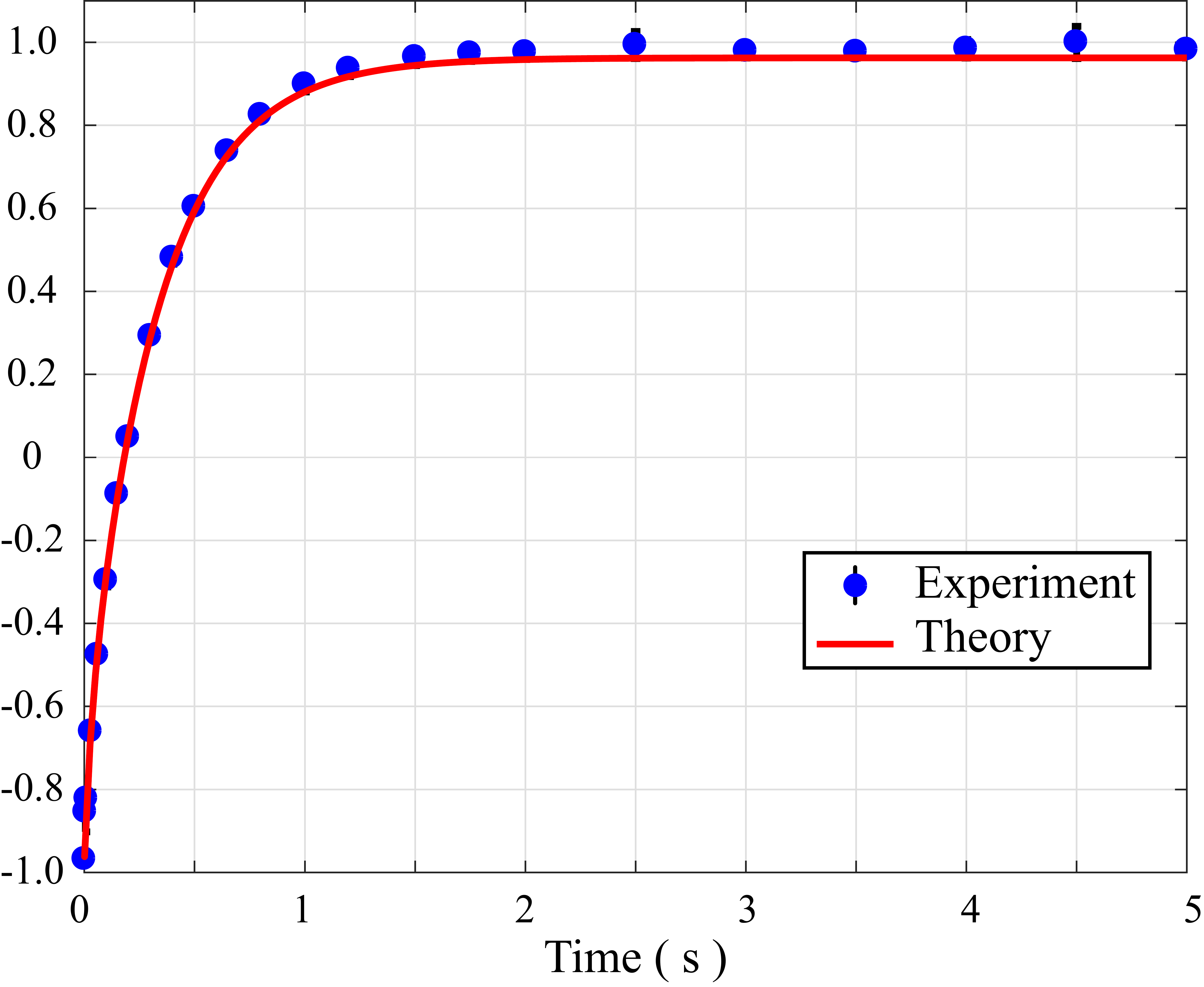} 
\caption[Longitudinal magnetization dynamic]{(Color online) Experimental data (blue dot symbols) {generated by } the inversion recovery pulse sequence, and the theoretical prediction (red solid line)  applying the Redfield's theory {of longitudinal magnetization} are displayed. {The relative error definition was applied to describe the error bars.} } \label{fig:MagnetizationZExpThe}
\end{figure}

{Considering} each density matrix element $\rho _{k,k+1}\left( t\right)$,  as denoted by the  Eq. (\ref{SolutionDensityMatrixElements}), and {each off-diagonal matrix element of the        $  {\mathbf{I}}_{x}  $  angular momentum operator, as denoted by  ${I}_{x}   \left( k,k+1\right) $, such that both   are used to   introduce the coefficients $A_{n}^{x}$ as} 
\begin{equation}
A_{n}^{x} = \sum_{p=1}^{p_{\text{max}}} 2 {I}_{x}   \left( p,p+1\right) \overline{W}_{n,p}^{   \left( 1\right)  }  \widetilde{\rho }_{n}\left( t_{0}\right) \text{.}
\end{equation}
{Consequently,} the mean value of the $x$-magnetization is
\begin{equation}
\left\langle {\mathbf{I}}_{x}\right\rangle = \mathcal{A}_{1}^{x} \left(   \sum_{n=1}^{n_{\text{max}}} A_{n}^{x} \exp \left[ -R_{n}^{\left( 1\right) }\left( t-t_{0}\right) \right]  \right)  \text{.} \label{ExponentialSuperpositionMagX}
\end{equation}
{Moreover, the}  initial condition {is represented by the } implementation of the initial quantum state. In this case, {the $\frac{\pi}{2}$-pulse efficiency of the}   Bloch-Echo  sequence is encoded by a second parameter  $ \mathcal{A}_{2}^{x} ${, and used as a proportionality coefficient between the}   theoretical initial state  ${\boldsymbol{\rho }}^{\text{theo}}\left( t_{0}\right)={\mathbf{I}}_{x}$ and the experimental initial state as denoted by $   {\boldsymbol{\rho }}^{\text{exp}}\left( t_{0}\right) =  \mathcal{A}_{2}^{x}  {\boldsymbol{\rho }}^{\text{theo}}\left( t_{0}\right)$.

\begin{figure}[t]
\includegraphics[width=3.40in]{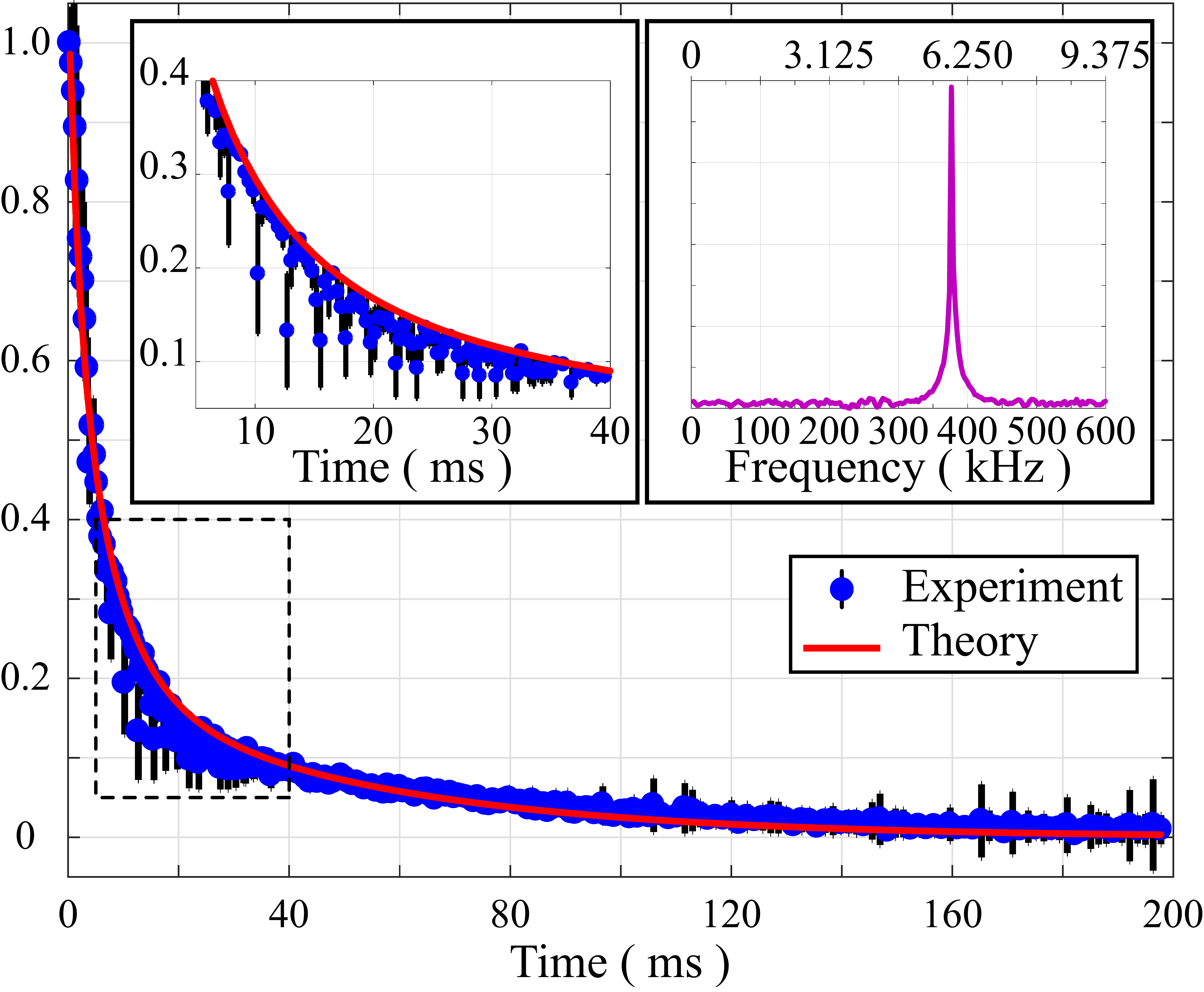} 
\caption[Transverse magnetization dynamic]{(Color online) Experimental data (blue dot symbols)  { generated by}  the Bloch-Echo pulse sequence, and the theoretical prediction (red solid line)  applying the Redfield's theory {of transverse magnetization} are displayed. {Insets: On the left, the dashed square region shows a zoom-in. On the right, a spectra calculated performing the fourier transform on the experimental data with characteristic frequency $\nu ^{\prime}= 376589$ Hz (or $\nu ^{*}= 5884$ Hz following the frequency axis on the top).  The relative error definition was applied to describe the error bars.    } } \label{fig:MagnetizationXExpThe}
\end{figure}

\begin{table}[b]
\begin{center}
\begin{ruledtabular}
\begin{tabular}{c|c c c|c}
Parameter  & Value $\pm$ Error & & Parameter & Value $\pm$ Error  \\   \cline{1-2}    \cline{4-5}
$ \mathcal{A}_{1}^{z} $	& $ 0.0230 \pm 0.0001$ & &	$ \mathcal{B}_{0}$ (\text{Hz}) & $ 83\pm 7$  \\
$ \mathcal{A}_{2}^{z} $	& $ 1.00   \pm 0.01  $ & &	$ \mathcal{B}_{1}$ (\text{Hz}) & $3.8\pm 0.3$ \\
$ \mathcal{A}_{1}^{x} $	& $ 0.019  \pm 0.001 $ & &	$ \mathcal{B}_{2}$ (\text{Hz}) & $0.18\pm 0.08$ \\
$ \mathcal{A}_{2}^{x} $	& $ 0.99   \pm 0.08  $ & &								   &				\\
\end{tabular} 
\end{ruledtabular}
\end{center} \caption{Parameter values generated by the fitting process comparing the theoretical equations and experimental data of longitudinal and transverse magnetization. $ \mathcal{A}_{1,2}^{x,z} $ are adimensional and $ \mathcal{B}_{0,1,2}$ are dimensional  fitting parameters.  } \label{tab:ParameterValues}
\end{table}

On the other hand, if the spectral density values $J_{k}$ with $k = 0,1,2$ are known then the solution of the linear system of differential equations can be found. In this sense and for practical {procedures on numerical optimization,}  we introduce the parameters $ \mathcal{B}_{k}$ with $k = 0,1,2$, which are proportional to each spectral density values  $ \mathcal{B}_{k} = C J_{k}$. 

Consequently,  Redfield's theory was used to establish a linear system of equations and the application of numerical procedures to optimize seven parameter values that allow finding an appropriate theoretical model that represents the experimental data.  

Basically, the computational task was performed using the Nelder–Mead method encoded in Matlab software. The execution of the numerical procedure generates an appropriate set of parameter values shown in Tab. \ref{tab:ParameterValues}.

{The $C$ parameter value is computed analyzing the}   Eq. (\ref{ConstantProportionalityGeneral}) and  {using}  the nuclear spin $I=7/2$, the quadrupole coupling definition $\frac{eQ}{\hbar }\mathcal{V}_{zz}\left( r\right) =\frac{4  I \left( 2I -1\right) \omega _{Q}}{6}$, the asymmetry parameter $\eta\approx 0$. Simplifying common constant coefficients, the parameter $C$ can be rewritten
\begin{equation}
C=\frac{\left(\omega _{Q} \right) ^{2}}{10} \text{.}
\end{equation}
From the spectra (see Fig. \ref{fig:SpectraSamplePFOCs}), the quadrupole angular frequency  is quantified $\omega _{Q} = 2 \pi \nu_{Q} = (2 \pi) 5969  $ \texttt{Hz}, such that the parameter is $C = 98.697 \times 10^{6} $ \texttt{Hz}$^{2}$. Therefore, using the values of Tab.  \ref{tab:ParameterValues} and the equivalent notation of the spectral density values  $ \mathcal{B}_{k} = C J_{k}$, they are described as follows
\begin{eqnarray}
 J_{0}  & = & \frac{ \mathcal{B}_{0}}{C} = (590 \pm 50) \times 10^{-9}  \texttt{s} \text{,} \\
 J_{1}  & = & \frac{ \mathcal{B}_{1}}{C} = (27 \pm 2) \times 10^{-9}  \texttt{s} \text{,} \\
 J_{2}  & = & \frac{ \mathcal{B}_{2}}{C} = (1.28 \pm 0.05) \times 10^{-9}  \texttt{s}\text{.}
 \end{eqnarray}

Therefore,  the parameter values {found and the magnetization theoretical prediction defined by Eq. (\ref{ExponentialSuperpositionMagZ}) and Eq. (\ref{ExponentialSuperpositionMagX}) are used to compute the time evolution and are displayed in Fig.  \ref{fig:MagnetizationZExpThe} and Fig. \ref{fig:MagnetizationXExpThe},}    respectively.

\begin{figure}[t]
\includegraphics[width=3.40in]{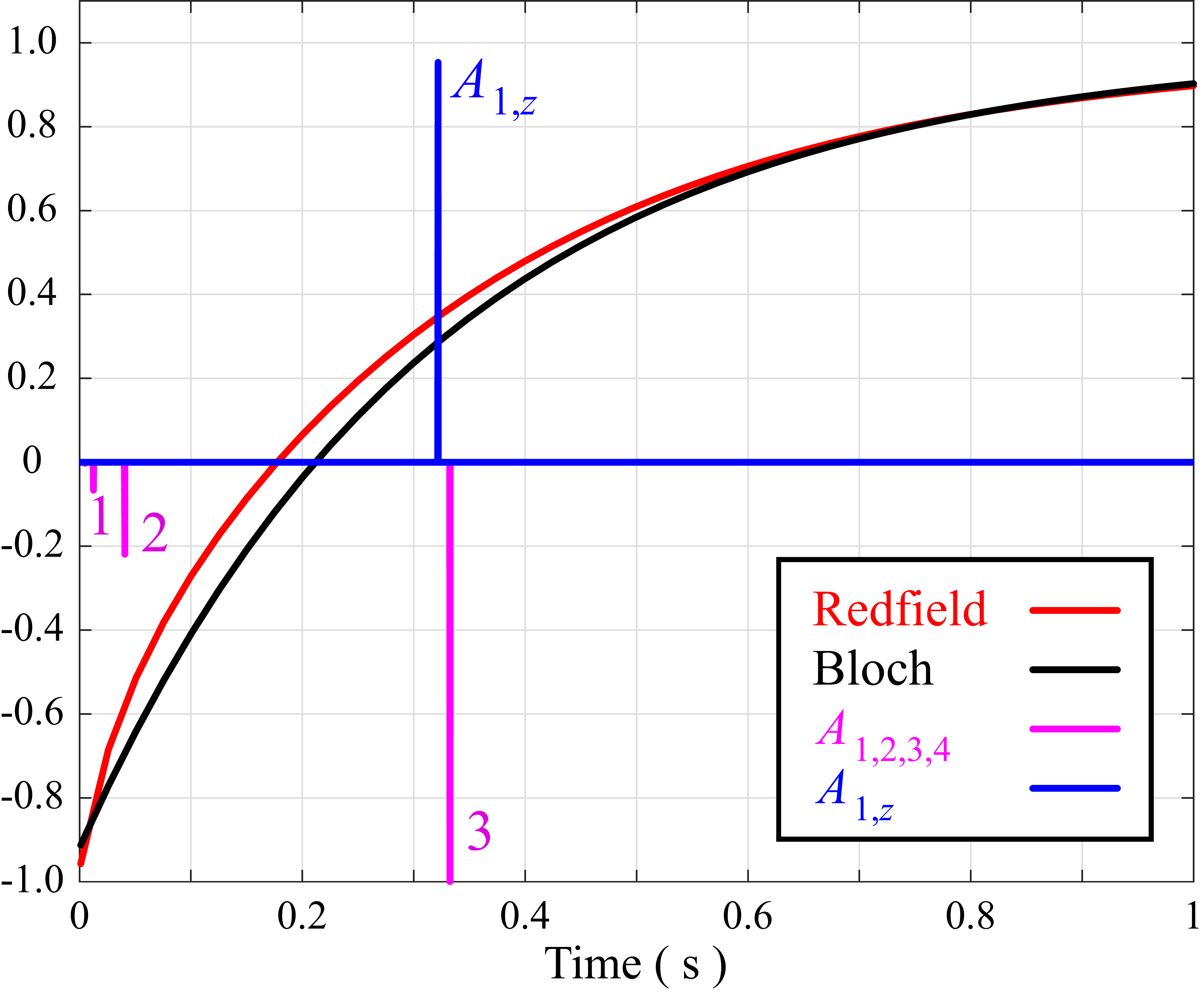} 
\caption[Comparison between Redfield's theory and Bloch decay model]{(Color online) The dynamics of the longitudinal magnetization predicted by Redfield's theory - $\left\langle {\mathbf{I}}_{z}\right\rangle$ (Bloch model - $\mathcal{M}_{z}$) is sketched by the solid red (black) line. Four (One) exponential functions are (is) used to generate the dynamics and they (it) are (is) sketched by the magenta (blue) solid vertical lines.}  \label{fig:ExponentialSuperpostionMagZ}
\end{figure} 

The application of  Redfield's theory to compute each element of the density matrix  generates  additional longitudinal time components,  T$_{k}^{\left( 0\right)}$. Each one contributes to the relaxation phenomena at different intensities, $A_{k}^{z}$, as summarized at the left side of Tab. \ref{tab:ExponentialSuperpostionMag}. From this data, the most important components that contribute effectively for the relaxation phenomena are  denoted with the subscripts $1$, $2$, $3$ and $4$, even though the $4$th-component contributes weakly; and the last four components do not contribute. Therefore, those quantification procedures allow us to identify four exponential functions capable to model the time evolution of the longitudinal magnetization.  A graphical representation of these parameter values is shown by the magenta vertical lines along a time axis, in a time window between 0 and 1 s, as sketched in Fig. \ref{fig:ExponentialSuperpostionMagZ}.

\begin{figure}[t]
\includegraphics[width=3.40in]{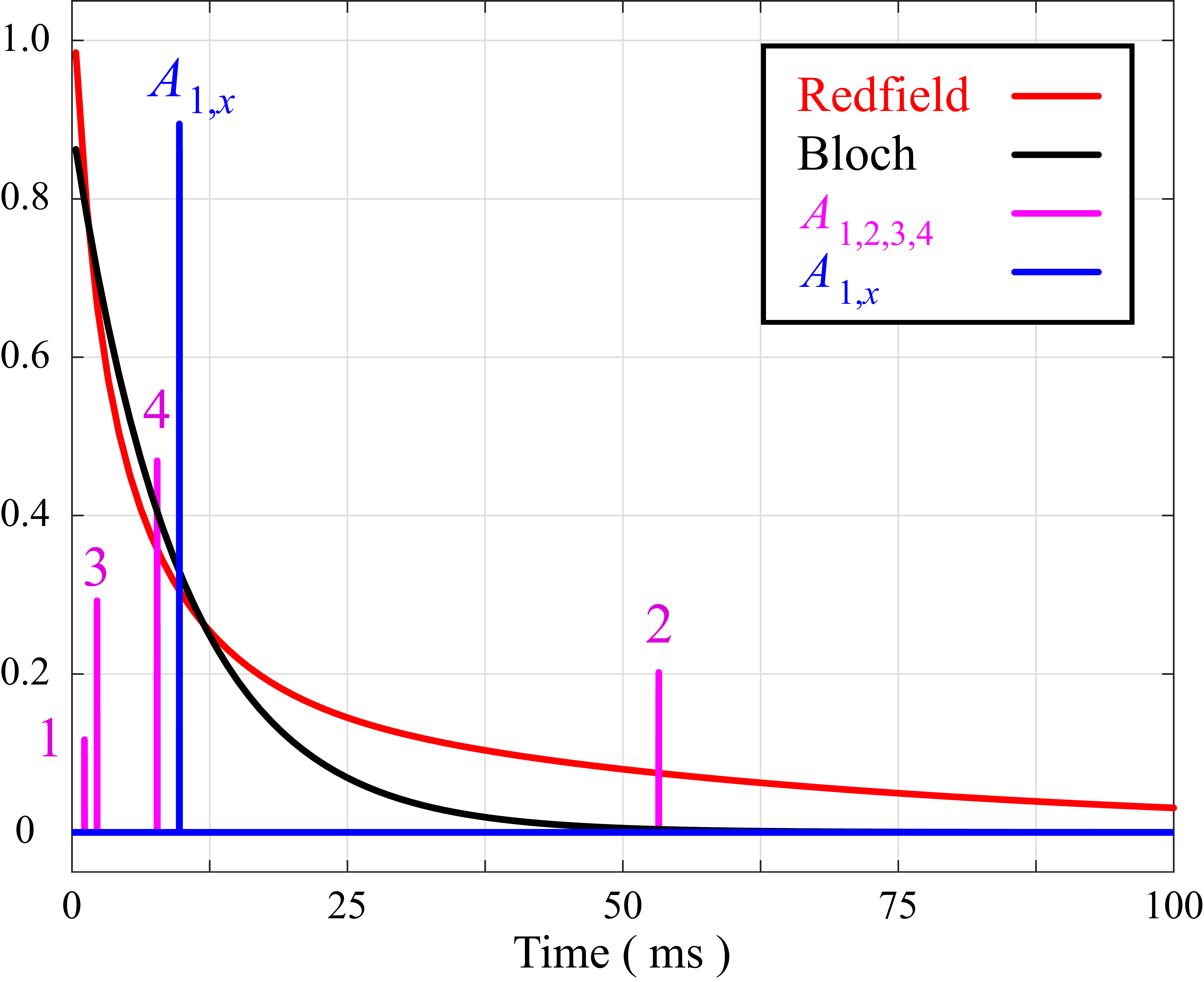} 
\caption[Comparison between Redfield's theory and Bloch decay model]{(Color online) The dynamics of the transverse magnetization predicted by Redfield's theory - $\left\langle {\mathbf{I}}_{x}\right\rangle$ (Bloch model - $\mathcal{M}_{x}$) is sketched by the solid red (black) line. Four (One) exponential functions are (is) used to generate the dynamics and they (it) are (is) sketched by the magenta (blue) solid vertical lines.}
\label{fig:ExponentialSuperpostionMagX}
\end{figure}

\begin{table}[b]
\begin{center}
\begin{ruledtabular}
\begin{tabular}{l c r c l c r}
\multicolumn{3}{c}{Longitudinal} & & \multicolumn{3}{c}{Transverse} \\      \cline{1-3}    \cline{5-7}
$k$		&     $A_{k}^{z} \times 10^{-3}$ & T$_{k}^{\left( 0\right)}$	(ms) & \qquad \qquad & $k$ 		&     $A_{k}^{x} \times 10^{-3}$ & T$_{k}^{\left( 1\right)}$	(ms)		\\     \cline{1-3}    \cline{5-7}
 $1$  & $ -67 \pm 1$  &  $11.5 \pm 0.7$ && $1$  & $111 \pm 9$  &  $1.15 \pm 0.08$   \\ 
 $2$  & $ -223 \pm 10$  &  $38 \pm 2$ && $2$  & $203 \pm 6$  &  $49 \pm 2$  \\ 
 $3$  & $-1672 \pm 8$ &  $310 \pm 10$ && $3$  & $292 \pm 9$  &  $2.3 \pm 0.2$  \\ 
 $4$  & $-2.6 \pm 0.1$  & $ 4.6 \pm 0.3$ && $4$  & $481 \pm 9$  &  $7.7 \pm 0.5$  \\ 
 $5$  & $0$  &  $4.6 \pm 0.3$  && $5$  & $0$  &  $1.15 \pm 0.08$ \\ 
 $6$  & $0$  &  $11.2 \pm 0.7$ && $6$  & $0$  &  $2.3 \pm 0.2$  \\ 
 $7$  & $0$  &  $37 \pm 2$	 && $7$  & $0$  &  $7.3 \pm 0.4$  \\ 
 $8$  & $0$  &  $\infty$     &&		 &		  &					\\
\end{tabular} 
\end{ruledtabular}
\end{center} \caption{On the left, amplitudes  $A_{k}^{z}$ and characteristic longitudinal times  T$_{k}^{\left( 0\right)}=1/R_{k}^{\left( 0\right) }$ used  to monitor the longitudinal magnetization represented by Eq. (\ref{ExponentialSuperpositionMagZ}), where the first four of them are sketched in Fig. \ref{fig:ExponentialSuperpostionMagZ}. On the right,  amplitudes   $A_{k}^{x}$ and characteristic transverse times  T$_{k}^{\left( 1\right)}=1/R_{k}^{\left( 1\right) }$ used to monitor the  transverse magnetization represented by  Eq. (\ref{ExponentialSuperpositionMagX}), where the first four of them are sketched in Fig. \ref{fig:ExponentialSuperpostionMagX}. } \label{tab:ExponentialSuperpostionMag}
\end{table}

An additional analysis is introduced by  using the traditional Bloch-decay method \cite{abragam1994Book}. 
 The longitudinal magnetization is described by a monoexponential model
\begin{equation}
\mathcal{M}_{z} = A_{0,z}+A_{1,z}   \left( 1- 2\exp\left[-t/T_{1} \right]  \right)   \text{,} \label{BlockModelT1} 
\end{equation}
where the parameters used to fit the experimental data have the following values: $A_{0,z}=0.04920 \pm 0.00002$,  $A_{1,z}=0.92450 \pm 0.00002$ and  $T_{1}=293.02 \pm 0.03$ ms, $T_{1}$ being the well known longitudinal relaxation time and it is displayed by the black solid line in Fig. \ref{fig:ExponentialSuperpostionMagZ}. Although, the $T_{1}$ value is different from any longitudinal characteristic time in Tab. \ref{tab:Js0maOrdem}, it  is very close to the $3$th-component T$_{3}^{\left( 0\right)}$.

Similarly, the transverse magnetization is modelled by the transverse time components, T$_{k}^{\left( 1\right)}$, at different intensity values,  $A_{k}^{x}$, as summarized at the right side of Tab.  \ref{tab:ExponentialSuperpostionMag}. From this data, the most important components that contribute to the relaxation phenomena are encoded by the first four components, such that the other ones have null amplitude values.    A graphical representation of these parameter values are represented by magenta vertical lines along of a time axis as sketched in Fig. \ref{fig:ExponentialSuperpostionMagX}.

Next, the transverse magnetization described by the Bloch-decay model \cite{abragam1994Book} follows a monoexponential time dependence and it is represented by the mathematical function
\begin{equation}
\mathcal{M}_{x} = A_{1,x}     \exp\left[-t/T_{2} \right]       \text{,} \label{BlockModelT2} 
\end{equation}
where the parameters used to fit the experimental data have following values:    $A_{1,x}=0.89450 \pm 0.00002$ and  $T_{2}= 9.7532 \pm 0.0005$ ms, with $T_{2}$ being the well known transverse  relaxation time and the time evolution of the transverse magnetization is displayed by the black solid line in Fig. \ref{fig:ExponentialSuperpostionMagZ}. This $T_{2}$ value is different, but close to the fourth transverse component T$_{4}^{\left( 1\right)}$ found at the right side of Tab. \ref{tab:ExponentialSuperpostionMag}, both   sketched in Fig. \ref{fig:ExponentialSuperpostionMagX}. {Another meaningful comment about the concave region (dashed square) of the experimental data on Fig. \ref{fig:MagnetizationXExpThe}, which seems very clustered, is zoomed and displayed at the left inset. It reveals some experimental data considerably far from the theoretical prediction, apparently happening in a periodic fashion. Thus, to verify our guess, the experimental data was analyzed by the Fourier transform. This procedure generates the spectrum shown at right inset. The single peak of the spectrum occurs at the frequency $\nu ^{\prime}= 376589$ Hz,  so dividing by an integer factor 64,  the  frequency value is $\nu ^{*}=5884$ Hz (see the horizontal axis on the top of the right inset). This value represents 1.4 \% of the quadrupolar frequency $\nu_{Q}$ and means an uncertainty on the experimental time step size. If the step size was more precise then the spectrum intensity of the inset would be lower.}

\begin{figure*}[t]
\begin{center}
    $\begin{array}{c@{\hspace{0.1in}}c}
     \multicolumn{1}{c}{\mbox{\bf }} & \multicolumn{1}{c}{\mbox{\bf }} \\ [-0.53cm]
     \epsfxsize=3.400in
     \epsffile{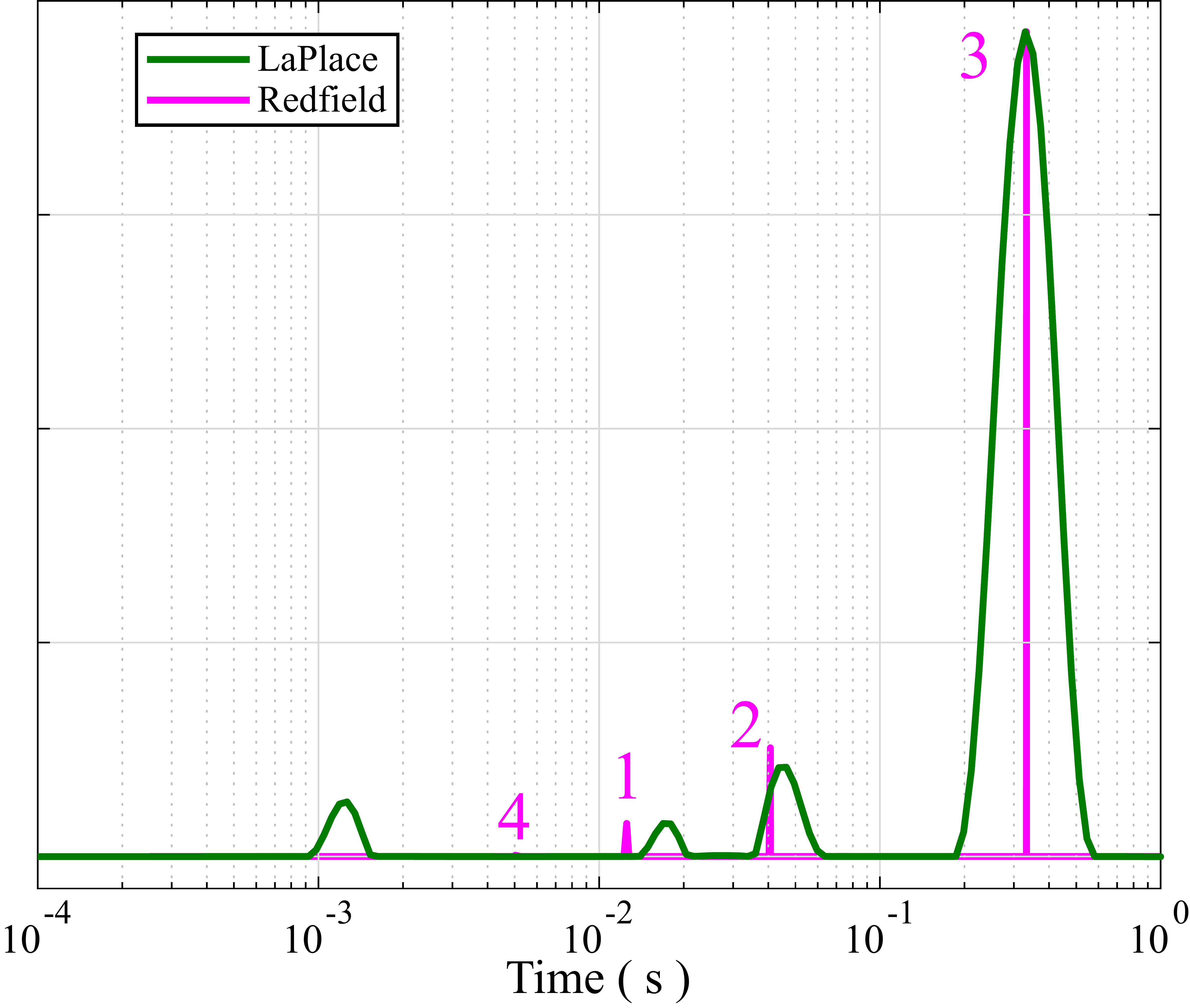} & 
     \epsfxsize=3.400in
     \epsffile{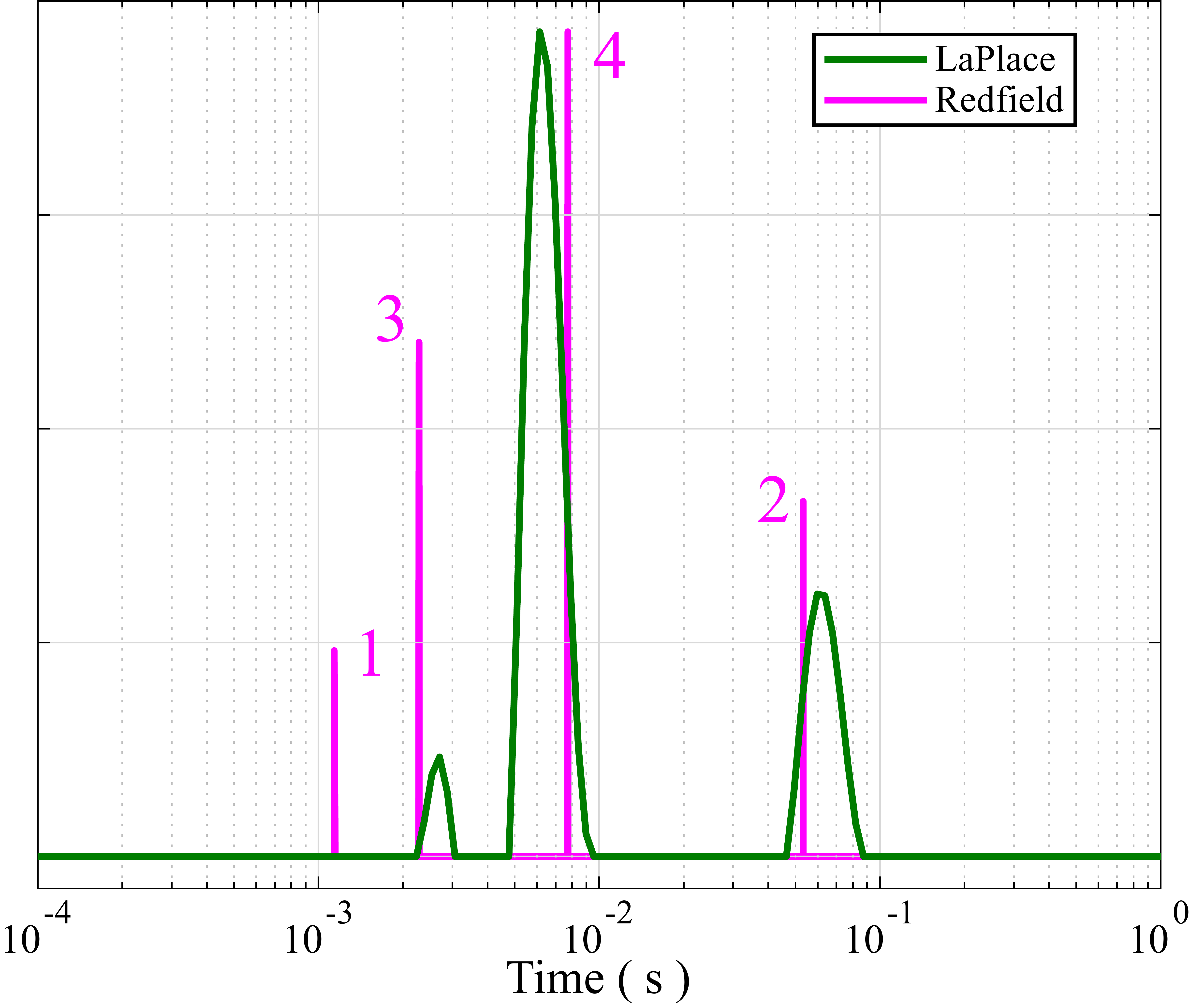} \\ [0.0cm] 
     \mbox{ \textbf{(a)} } & \mbox{ \textbf{(b)} } 
     \end{array}$
\end{center}
\caption{(Color online) Comparison of the characteristic time computed by Redfield's theory (solid magenta lines) and distribution time computed by a numerical procedure of inverse Laplace transformation tool (solid dark green lines). 	(a)  The longitudinal characteristic time. (b) The transverse characteristic time. } \label{fig:MagComparaLaPlaceRedfield}
\end{figure*}
Also, the longitudinal and transverse magnetization experimental data, Fig. \ref{fig:MagnetizationZExpThe} and Fig. \ref{fig:MagnetizationXExpThe}, were  analyzed using an inverse Laplace transformation (ILT) tool, frequently used in relaxometry studies \cite{TRANSFOR, Marcel}. The characteristic time distributions generated by the ILT are displayed in Fig. \ref{fig:MagComparaLaPlaceRedfield} and compared to the time constants achieved with Redfield's theory. 
The longest and most intense components are well predicted by the ILT and Redfield theory, but the correlation between the methods decreases for components that have lower intensity and shorter times. There are different aspects that can explain the difference between the times obtained. However, the correlation obtained, mainly for the $T_{1}$ relaxation times, demonstrates the versatility in determining the system's relaxation times from the Redfield’s method. The transverse relaxation time is strongly influenced by the inhomogeneous field and the diffusion process, as well as by the imprecision of the RF pulses. This could be one of the explanations for the difference between the methods for transverse relaxation. Therefore, the numerical fitting procedure to extract information of the spin system using Redfield's theory generates acceptable accuracy predictions.

\section{Conclusions}
\label{sec:Conclusions} 

The relaxation study of a nuclear spin system with $I=7/2$ using Redfield's master equation is developed in this paper. Analytical solutions of the density matrix elements are introduced and confirmed by the multiexponential description of each of them, save the highest-order element. 

The number of characteristic relaxation time components used to describe the relaxation phenomena for the $q$ coherence order density matrix element with spin value $I$ is computed by $n_{I}=2I+1-q$. 

The theoretical solutions previously reported on the dynamic evolution of the spin system under the four exponential dependence of the longitudinal and transverse magnetization are confirmed using experimental data of $^{133}$Cs signals of a Cs-PFO lyotropic liquid crystal sample.  From a physical point of view, the density matrix elements solutions open new possibilities to explore properties of open quantum systems and master equations on quantum information processing and solid state matter spectroscopy.

The methodology on analyzing the monotonic decay of longitudinal and transverse magnetization  seems a promissory procedure for spectroscopy, even to introduce and characterize new fluctuation functions of any  thermal bath coupled weakly with the spin system.

\section{Acknowledgements}

The authors  acknowledge  CNPq INCT-QI (465469/2014-0). A.C.S.L. acknowledges CNPq (142118/2018-4).  T.J.B. acknowledges financial support from CNPq (308076/2018-4) and FAPESP (2012/02208-5). R.A. acknowledges CNPq (309023/2014-9, 459134/2014-0). This study was financed in part by the Coordena\c{c}\~{a}o de Aperfei\c{c}oamento de Pessoal de N\'{i}vel Superior - Brasil (CAPES) - Finance Code 001. We also acknowledge Patrick Judeinstein for the Cs-PFO samples.


\appendix

\section{Solving the Redfield equation at each coherence order}
\label{app:SolvingRedfieldEquation}

\textit{Seventh coherence order:} To describe the main algebraic procedures and calculus definitions appropriately, let's apply Redfield's theory to compute the solution of the density matrix elements of $7$th order.  One of the main information that Redfield's equation provides us is a set of differential equations. For this initial step, {Redfield's equation denoted by Eq. (\ref{RelaxationDensityMatrixElementsCorrigida})  and the 2nd-rank tensor operators of Eq. (\ref{QuadrupolarBathSecondMinus7o2}-\ref{QuadrupolarBathZero7o2}) are analyzed assuming ${\mathbf{I}}_{r}=  {\mathbf{T}}_{7,-7} $. Applying some algebraic procedures, } it generates one differential equation as denoted by
\begin{equation}
\frac{1}{36C}\frac{d \rho _{8,1} }{dt} = \boldsymbol{\mathcal{J}}^{\left( 7\right) } \rho _{8,1}\text{.} \label{EqDiOrRank7Order7Spin7o2Right}
\end{equation}
where the seventh order superoperator  is denoted by $\boldsymbol{\mathcal{J}}^{\left( 7\right) }=-21J_{1}-7J_{2}$. By the mathematical procedures to solve differential equations, the solution for the density matrix element is denoted as
\begin{equation*}
\rho _{8,1}\left( t\right)  =\exp \left[ -R_{1}^{\left( 7\right) }\left( t-t_{0}\right) %
\right] \rho _{8,1}\left( t_{0}\right) \text{,}
\end{equation*}%
where the relaxation rate of $7$th order is resumed in Tab.  \ref{tab:Eigenvalues}. The relaxation rate $R_{1}^{\left( 7\right) }$ matches with a previous study assessing multiple quantum coherences of $Cs+$ bound to a crown ether ($18$-crown-$6$). The relaxation rate $R_{1}^{(7)}$ is proportional to the relaxation rate of Eq. (A9) of Ref.\cite{tsoref1996}.

\textit{Sixth coherence order:}  Redfield's theory performed to compute the density matrix elements of $6$th order are  denoted by $\rho _{7,1}\left( t\right)$  and $\rho _{8,2}\left( t\right) $. {To achieve this task, Redfield's equation denoted by  Eq. (\ref{RelaxationDensityMatrixElementsCorrigida}) and the 2nd-rank tensor operator of Eq. (\ref{QuadrupolarBathSecondMinus7o2}-\ref{QuadrupolarBathZero7o2}) are analyzed assuming  ${\mathbf{I}}_{r}=  {\mathbf{T}}_{6,-6} $ and  ${\mathbf{I}}_{r}=  {\mathbf{T}}_{7,-6} $.  } The execution of mathematical procedures as discussed some paragraphs above generates a linear system of differential equations, it can be represented using  matrix notation
\begin{equation*}
\frac{1}{36C}\left[ 
\begin{array}{c}
\frac{d\rho _{7,1}}{dt} \\ 
\frac{d\rho _{8,2}}{dt}%
\end{array}%
\right] = \boldsymbol{\mathcal{J}}^{\left( 6\right) } \left[ 
\begin{array}{c}
\rho _{7,1} \\ 
\rho _{8,2}%
\end{array}%
\right] \text{,}
\end{equation*}
where the sixth order superoperator is denoted by
\begin{equation*}
\boldsymbol{\mathcal{J}}^{\left( 6\right) }  =\left[ 
\begin{array}{cc}
-9J_{0}-29J_{1}-11J_{2} & -21J_{1} \\ 
-21J_{1} & -9J_{0}-29J_{1}-11J_{2}%
\end{array}%
\right] \text{.}
\end{equation*}
Next, by performing the fundamentals on differential equations is needed to find a transformation matrix $\mathbf{W}^{ \left( 6 \right)}$ that diagonalizes the superoperator and also must satisfy the property $\mathbf{W}^{ \left( 6 \right)}\overline{\mathbf{W}}^{ \left( 6 \right)}=\overline{\mathbf{W}}^{ \left( 6 \right)}{\mathbf{W}}^{ \left( 6 \right)}=\mathbb{I}$, where ${\mathbb{I}}$ denotes the identity matrix.  In this case, the transformation ${\mathbf{W}}^{ \left( 6 \right)}$ is denoted matricially by
\begin{equation*}
{\mathbf{W}}^{ \left( 6 \right)}=\left[ 
\begin{array}{cc}
W_{1,1}^{ \left( 6 \right)} & W_{1,2}^{ \left( 6 \right)}\\ 
W_{2,1}^{ \left( 6 \right)} & W_{2,2}^{ \left( 6 \right)}
\end{array}%
\right] =\frac{1}{\sqrt{2}}\left[ 
\begin{array}{cc}
-1 & 1 \\ 
1 & 1%
\end{array}%
\right] \text{,}
\end{equation*}
and the inverse transformation  $\overline{\mathbf{W}}^{ \left( 6 \right)}$ using the matrix notation is represented by
\begin{equation*}
\overline{\mathbf{W}}^{ \left( 6 \right)}=\left[ 
\begin{array}{cc}
\overline{W}_{1,1}^{  \left( 6 \right)} & \overline{W}_{1,2}^{  \left( 6 \right)} \\ 
\overline{W}_{2,1}^{  \left( 6 \right)} & \overline{W}_{2,2}^{ \left( 6 \right)}
\end{array}%
\right] =\frac{1}{\sqrt{2}}\left[ 
\begin{array}{cc}
-1 & 1 \\ 
1 & 1%
\end{array}%
\right] \text{.}
\end{equation*}
Therefore,  the solutions are written in an extended form
\begin{widetext}
\begin{eqnarray*}
\rho _{7,1}\left( t\right)  &=&\overline{W}_{1,1}^{  \left( 6 \right)} \exp \left[ -R_{1}^{\left(
6\right) }\left( t-t_{0}\right) \right] \widetilde{\rho }_{1}\left(
t_{0}\right) +\overline{W}_{1,2}^{  \left( 6 \right)} \exp \left[ -R_{2}^{\left( 6\right) }\left(
t-t_{0}\right) \right] \widetilde{\rho }_{2}\left( t_{0}\right) \text{,} \\
\rho _{8,2}\left( t\right)  &=&\overline{W}_{2,1}^{  \left( 6 \right)} \exp \left[ -R_{1}^{\left(
6\right) }\left( t-t_{0}\right) \right] \widetilde{\rho }_{1}\left(
t_{0}\right) +\overline{W}_{2,2}^{  \left( 6 \right)} \exp \left[ -R_{2}^{\left( 6\right) }\left(
t-t_{0}\right) \right] \widetilde{\rho }_{2}\left( t_{0}\right) \text{,}
\end{eqnarray*}
\end{widetext}
where the relaxation rates of $6$th order are resumed the eigenvalue interpretation in Tab.  \ref{tab:Eigenvalues}. Also, the relaxation rate $R_{1}^{\left( 6\right) }$ is proportional to one other and denoted by Eq. (A8) in Ref. \cite{tsoref1996}.

Let us introduce a general notation to represent the solutions for any density matrix element.  Considering a subscript index $p \equiv \left\{ 1 , 2\right\}$ to represent the number of solutions of the linear system, such that it matches with the index used to represent the columns of the transformation. Therefore, the solution can be rewritten as 
\begin{eqnarray}
\rho _{7,1}\left( t\right)  &=&\sum_{p=1}^{2} \overline{W}_{1,p}^{  \left( 6 \right)} \exp \left[ -R_{p}^{\left(
6\right) }\left( t-t_{0}\right) \right] \widetilde{\rho }_{p}\left(
t_{0}\right)  \text{,} \label{SolutionRho7o1} \\
\rho _{8,2}\left( t\right)  &=&\sum_{p=1}^{2} \overline{W}_{2,p}^{  \left( 6 \right)} \exp \left[ -R_{p}^{\left(
6\right) }\left( t-t_{0}\right) \right] \widetilde{\rho }_{p}\left(
t_{0}\right)  \text{.} \label{SolutionRho8o2}
\end{eqnarray}
Also, considering another subscript index  $n \equiv \left\{ 1 , 2\right\}$ to represent the number of linear equations, which matches with the subscript index of the transformation row number. Both subscript indices  can be bounded using a practical rule that depends on the spin value $I$ and the coherence order value $q$ such that they are denoted by  $p_{\text{max}},n_{\text{max}}=2I+1-q$. In this case,  coherence order is  $q=6$  then the linear system can be simplified by the mathematical expression
\begin{equation}
\rho _{q+n,n}\left( t\right)   = \sum_{p=1}^{p_{\text{max}}} \overline{W}_{n,p}^{  \left(q \right)} \exp \left[ -R_{p}^{\left(
q\right) }\left( t-t_{0}\right) \right] \widetilde{\rho }_{p}\left( t_{0}\right)  \text{,} \label{RedfieldElementoGeral}
\end{equation}
and this expression will be very useful to represent any density matrix element below and even the main diagonal elements. Another important definition is about the initial conditions $ \widetilde{\rho }_{p}\left( t_{0}\right)$ represented at their matrix notation 
\begin{equation*}
 \left[ 
\begin{array}{c}
 \widetilde{\rho }_{1}\left( t_{0}\right)   \\ 
 \widetilde{\rho }_{2}\left( t_{0}\right)  
\end{array}%
\right]=\left[ 
\begin{array}{cc}
W_{1,1}^{  \left( 6 \right)}  & W_{1,2}^{  \left( 6 \right)}  \\ 
W_{2,1}^{  \left( 6 \right)}  & W_{2,2}^{  \left( 6 \right)} 
\end{array}%
\right] \left[ 
\begin{array}{ c}
  \rho _{7,1}\left( t_{0}\right) \\ 
  \rho _{8,2}\left( t_{0}\right)
\end{array}%
\right] \text{,}
\end{equation*}
where the initial parameters depend on the density matrix elements of the initial quantum state. The initial parameters for any coherence order
\begin{equation}
 \widetilde{\rho }_{p}\left( t_{0}\right)   = \sum_{n=1}^{n_{\text{max}}} W_{p,n}^{  \left( q \right)}  \ \rho _{q+n,n}\left( t_{0}\right)   \text{,}  \label{RedfieldElementoGeralInitialConditions}
\end{equation}
therefore, the most interesting equations in this generalization procedure are the Eqs. (\ref{RedfieldElementoGeral}) and  (\ref{RedfieldElementoGeralInitialConditions}). These both equations will be applied to other set of density matrix elements; consequently, if the transformation $\mathbf{W}^{  \left( 6 \right)} $ and $\mathbf{W}^{ \prime \left( 6 \right)} $ are well established then the density matrix elements will be well defined.  

\begin{widetext}
\textit{Fifth coherence order:} The density matrix elements of $5$th order with the elements of the density matrix denoted as $\rho _{6,1}\left( t\right) $, $\rho _{7,2}\left( t\right) $ and $\rho _{8,3}\left( t\right) $ can be represented with the Eq. (\ref{RedfieldElementoGeral}) and (\ref{RedfieldElementoGeralInitialConditions}) such that the subscript indices $p \equiv \left\{ 1 , 2 , 3\right\}$,  $n \equiv \left\{ 1 , 2 , 3\right\}$ and the coherence order $q=5$. With this notation only the transformation matrix must be resumed at its extended notation. Therefore, the transformation $\mathbf{W}^{  \left( 5\right)} $ is denoted explicitly by
\begin{equation*}
\mathbf{W}^{  \left( 5\right)}=\left[ 
\begin{array}{ccc}
W_{1,1}^{  \left( 5\right)} & W_{1,2}^{  \left( 5\right)} & W_{1,3}^{  \left( 5\right)} \\ 
W_{2,1}^{  \left( 5\right)} & W_{2,2}^{  \left( 5\right)} & W_{2,3}^{  \left( 5\right)} \\ 
W_{3,1}^{  \left( 5\right)} & W_{3,2}^{  \left( 5\right)} & W_{3,3}^{  \left( 5\right)}
\end{array}%
\right] =\frac{1}{\sqrt{26}}\left[ 
\begin{array}{ccc}
\frac{\sqrt{6}-\sqrt{7}b_{1}}{\left( a_{1}-b_{1}\right) }\sqrt{a_{1}^{2}+1}
& -\frac{\sqrt{12}b_{1}+\sqrt{14}}{\left( a_{1}-b_{1}\right) }\sqrt{%
a_{1}^{2}+1} & \frac{\sqrt{6}-\sqrt{7}b_{1}}{\left( a_{1}-b_{1}\right) }%
\sqrt{a_{1}^{2}+1} \\ 
-\sqrt{13} & 0 & \sqrt{13} \\ 
-\frac{\sqrt{6}-\sqrt{7}a_{1}}{\left( a_{1}-b_{1}\right) }\sqrt{b_{1}^{2}+1}
& \frac{\sqrt{12}a_{1}+\sqrt{14}}{\left( a_{1}-b_{1}\right) }\sqrt{%
b_{1}^{2}+1} & -\frac{\sqrt{6}-\sqrt{7}a_{1}}{\left( a_{1}-b_{1}\right) }%
\sqrt{b_{1}^{2}+1}%
\end{array}%
\right] \text{,}
\end{equation*}
and the inverse transformation  denoted by $\overline{\mathbf{W}}^{  \left( 5\right)}$ is
\begin{equation*}
\overline{\mathbf{W}}^{  \left( 5\right)}=\left[ 
\begin{array}{ccc}
\overline{W}_{1,1}^{\left( 5\right)} & \overline{W}_{1,2}^{  \left( 5\right)} & \overline{W}_{1,3}^{  \left( 5\right)} \\ 
\overline{W}_{2,1}^{  \left( 5\right)} & \overline{W}_{2,2}^{ \left( 5\right)} & \overline{W}_{2,3}^{ \left( 5\right)} \\ 
\overline{W}_{3,1}^{  \left( 5\right)} & \overline{W}_{3,2}^{  \left( 5\right)} & \overline{W}_{3,3}^{  \left( 5\right)}
\end{array}%
\right] =\frac{1}{\sqrt{26}}\left[ 
\begin{array}{ccc}
\frac{\sqrt{7}+\sqrt{6}a_{1}}{\sqrt{a_{1}^{2}+1}} & -\sqrt{13} & \frac{\sqrt{%
7}+\sqrt{6}b_{1}}{\sqrt{b_{1}^{2}+1}} \\ 
\frac{\sqrt{12}-\sqrt{14}a_{1}}{\sqrt{a_{1}^{2}+1}} & 0 & \frac{\sqrt{12}-%
\sqrt{14}b_{1}}{\sqrt{b_{1}^{2}+1}} \\ 
\frac{\sqrt{7}+\sqrt{6}a_{1}}{\sqrt{a_{1}^{2}+1}} & \sqrt{13} & \frac{\sqrt{7%
}+\sqrt{6}b_{1}}{\sqrt{b_{1}^{2}+1}}%
\end{array}%
\right] \text{,}
\end{equation*}
where the constant coefficients are denoted by
\begin{eqnarray*}
a_{1} &=&\frac{25J_{0}+656J_{1}-5J_{2}-13\sqrt{ 625J_{0}^{2} - 800J_{0}J_{1} - 250J_{0}J_{2} +2944J_{1}^{2} +160J_{1}J_{2}+25J_{2}^{2} }}{10\sqrt{42}\left( -5J_{0}+4J_{1}+J_{2}\right) }\text{,} \\
b_{1} &=&\frac{25J_{0}+656J_{1}-5J_{2}+13\sqrt{ 625J_{0}^{2} - 800J_{0}J_{1} - 250J_{0}J_{2} +2944J_{1}^{2} +160J_{1}J_{2}+25J_{2}^{2} }}{10\sqrt{42}\left( -5J_{0}+4J_{1}+J_{2}\right) }\text{,}
\end{eqnarray*}
and following the general notation of Eq. (\ref{RedfieldElementoGeral}), the fifth order density matrix elements are denoted by
\begin{eqnarray*}
\rho _{6,1}\left( t\right)  &=&\overline{W}_{1,1}^{ \left( 5\right)}\exp \left[ -R_{1}^{\left(
5\right) }\left( t-t_{0}\right) \right] \widetilde{\rho }_{1}\left(
t_{0}\right) +\overline{W}_{1,2}^{ \left( 5\right) }\exp \left[ -R_{2}^{\left( 5\right) }\left(
t-t_{0}\right) \right] \widetilde{\rho }_{2}\left( t_{0}\right)
+\overline{W}_{1,3}^{ \left( 5\right) }\exp \left[ -R_{3}^{\left( 5\right) }\left( t-t_{0}\right) %
\right] \widetilde{\rho }_{3}\left( t_{0}\right) \text{,} \\
\rho _{7,2}\left( t\right)  &=&\overline{W}_{2,1}^{ \left( 5\right) }\exp \left[ -R_{1}^{\left(
5\right) }\left( t-t_{0}\right) \right] \widetilde{\rho }_{1}\left(
t_{0}\right) +\overline{W}_{2,3}^{ \left( 5\right) }\exp \left[ -R_{3}^{\left( 5\right) }\left(
t-t_{0}\right) \right] \widetilde{\rho }_{3}\left( t_{0}\right) \text{,} \\
\rho _{8,3}\left( t\right)  &=&\overline{W}_{3,1}^{ \left( 5\right) }\exp \left[ -R_{1}^{\left(
5\right) }\left( t-t_{0}\right) \right] \widetilde{\rho }_{1}\left(
t_{0}\right) +\overline{W}_{3,2}^{ \left( 5\right)}\exp \left[ -R_{2}^{\left( 5\right) }\left(
t-t_{0}\right) \right] \widetilde{\rho }_{2}\left( t_{0}\right)
+\overline{W}_{3,3}^{ \left( 5\right)}\exp \left[ -R_{3}^{\left( 5\right) }\left( t-t_{0}\right) %
\right] \widetilde{\rho }_{3}\left( t_{0}\right) \text{,}
\end{eqnarray*}
where the relaxation rates of $5$th order are resumed at their eigenvalue notation in Tab.  \ref{tab:Eigenvalues}. Also, the relaxation rate $R_{1}^{\left( 5\right) }$ and $R_{3}^{\left( 5\right) }$   are analytical and compatible solutions of the relaxation operator denoted at  Eq. (A7a) of Ref. \cite{tsoref1996}.

\begin{table}
\begin{center}
\begin{tabular}{|c|c|c|} \hline \hline
$n$	& $k$	& Eigenvalues $ \lambda _{k}^{\left( n\right) }$  of Redfield's superoperator $\boldsymbol{\mathcal{J}}^{\left( n\right) }$     	\\ \hline \hline
7 & $1$ &      $-\left(21J_{1}+7 J_{2} \right)$ 		\\ \hline
6 & $1$	& $-\left( 9J_{0}+8J_{1}+11J_{2} \right)$	\\
  & $2$	& $-\left( 9J_{0}+50J_{1}+11J_{2} \right)$	\\ \hline
5 & $1$	& $-\frac{25}{2}J_{0}-29J_{1}-\frac{25}{2} J_{2}-\frac{1}{2}\sqrt{ 625J_{0}^{2}-800J_{0}J_{1}-250J_{0}J_{2}+2944J_{1}^{2}+160J_{1}J_{2}+25J_{2}^{2} }$		\\
  & $2$	& $ -25J_{0}-21J_{1}-24J_{2} $	\\
  & $3$	& $ -\frac{25}{2}J_{0}-29J_{1}-\frac{25}{2} J_{2}+\frac{1}{2}\sqrt{ 625J_{0}^{2}-800J_{0}J_{1}-250J_{0}J_{2}+2944J_{1}^{2}+160J_{1}J_{2}+25J_{2}^{2} } $	\\ \hline
4 & $1$	& $ -20J_{0}-29J_{1}-21J_{2}-\sqrt{256\left( J_{0}-J_{1}\right)^{2}+105\left( J_{1}-J_{2}\right) ^{2}} $	\\
  & $2$	& $ -20J_{0}-29J_{1}-21J_{2}+\sqrt{256\left( J_{0}-J_{1}\right)^{2}+105\left( J_{1}-J_{2}\right) ^{2}} $	\\
  & $3$	& $ -20J_{0}-13J_{1}-21J_{2}-\sqrt{256J_{0}^{2}+105\left( J_{1}+J_{2}\right) ^{2}} $	\\
  & $4$	& $ -20J_{0}-13J_{1}-21J_{2}+\sqrt{256J_{0}^{2}+105\left( J_{1}+J_{2}\right) ^{2}} $	\\ \hline
3 & $1$	& $  \iota _{1}^{\left( 3\right) } \left( \alpha,\beta, \gamma, \lambda _{3}^{\left( 3\right) }  \right) $	\\ 
  & $2$	& $ -9J_{0}-21J_{1}-40J_{2} $	\\ 
  & $3$	& $  \lambda _{3}^{\left( 3\right) } \left( \alpha ,\beta , \gamma   \right)  $	\\ 
  & $4$	& $ -36J_{0}-13J_{1}-21J_{2} $	\\ 
  & $5$	& $   \varsigma _{5}^{\left( 3\right) } \left( \alpha,\beta, \gamma, \lambda _{3}^{\left( 3\right) }  \right) $	\\ \cline{2-3}
  & \multicolumn{2}{|l|}{ $ \gamma=-6804J_{0}^{2}J_{1}-8748J_{0}^{2}J_{2}-12573J_{0}J_{1}^{2}-35100J_{0}J_{1}J_{2}-12\,303J_{0}J_{2}^{2}-3653J_{1}^{3}-16002J_{1}^{2}J_{2}-13947J_{1}J_{2}^{2}-2870J_{2}^{3}\text{,}$}	\\
  & \multicolumn{2}{|l|}{ $ \beta=324J_{0}^{2}+1818J_{0}J_{1}+1764J_{0}J_{2}+827J_{1}^{2}+2140J_{1}J_{2}+767J_{2}^{2}\text{,} $ }	\\
  & \multicolumn{2}{|l|}{ $ \alpha=-45J_{0}-55J_{1}-58J_{2}\text{.} $ }	\\ \hline
2 & $1$	& $  \iota _{1}^{\left( 2\right) } \left( \alpha _{\mathcal{A}} ,\beta _{\mathcal{A}}, \gamma _{\mathcal{A}}, \lambda _{2}^{\left( 2\right) }  \right) $	\\
  & $2$	& $ \lambda _{2}^{\left( 2\right) } \left( \alpha _{\mathcal{A}} ,\beta _{\mathcal{A}}, \gamma _{\mathcal{A}}  \right) $	\\
  & $3$	& $ \iota _{3}^{\left( 2\right) } \left( \alpha _{\mathcal{B}} ,\beta _{\mathcal{B}}, \gamma _{\mathcal{B}}, \lambda _{4}^{\left( 2\right) }  \right) $	\\
  & $4$	& $ \lambda _{4}^{\left( 2\right) } \left( \alpha _{\mathcal{B}} ,\beta _{\mathcal{B}}, \gamma _{\mathcal{B}}  \right) $	\\
  & $5$	& $ \varsigma _{5}^{\left( 2\right) } \left( \alpha _{\mathcal{A}} ,\beta _{\mathcal{A}}, \gamma _{\mathcal{A}}, \lambda _{2}^{\left( 2\right) }  \right) $	\\
  & $6$	& $ \varsigma _{6}^{\left( 2\right) } \left( \alpha _{\mathcal{B}} ,\beta _{\mathcal{B}}, \gamma _{\mathcal{B}}, \lambda _{4}^{\left( 2\right) }  \right) $	\\ \cline{2-3}
  & \multicolumn{2}{|l|}{$ \gamma _{\mathcal{A}}=-225J_{0}^{3}-4764J_{0}^{2}J_{1}-7753J_{0}^{2}J_{2}-13188J_{0}J_{1}^{2}-37020J_{0}J_{1}J_{2}-15%
783J_{0}J_{2}^{2}-6048J_{1}^{3}-26292J_{1}^{2}J_{2}-21552J_{1}J_{2}^{2}-4575J_{2}^{3} \text{,}$} \\
  & \multicolumn{2}{|l|}{$ \beta _{\mathcal{A}}=259J_{0}^{2}+1368J_{0}J_{1}+1874J_{0}J_{2}+1092J_{1}^{2}+2940J_{1}J_{2}+1287J_{2}^{2}\text{,}$} \\
  & \multicolumn{2}{|l|}{$ \alpha _{\mathcal{A}}=-35J_{0}-60J_{1}-73J_{2}\text{,}$} \\
  & \multicolumn{2}{|l|}{$ \gamma _{\mathcal{B}}=-225J_{0}^{3}-2514J_{0}^{2}J_{1}-7753J_{0}^{2}J_{2}-6048J_{0}J_{1}^{2}-29240J_{0}J_{1}J_{2}-15783J_{0}J_{2}^{2}-2688J_{1}^{3}-17472J_{1}^{2}J_{2}-25702J_{1}J_{2}^{2}-4575J_{2}^{3}%
\text{,}$}	\\
  & \multicolumn{2}{|l|}{$ \beta _{\mathcal{B}}=259J_{0}^{2}+1028J_{0}J_{1}+1874J_{0}J_{2}+672J_{1}^{2}+2520J_{1}J_{2}+1287J_{2}^{2}\text{,}$}	\\
  & \multicolumn{2}{|l|}{$ \alpha _{\mathcal{B}}=-35J_{0}-50J_{1}-73J_{2}\text{.}$}	\\ \hline \hline
\end{tabular}
\end{center}
\caption{The list of eigenvalues of each Redfield's superoperator $\boldsymbol{\mathcal{J}}^{\left( n\right)}$ for the spin $I=7/2$ with $n$ means the coherence order, $k$ means an index to distinguish the eigenvalue at each order.  Some eigenvalues depend on the mathematical expression $ \lambda _{k}^{\left( n\right) } \left( \alpha ,\beta, \gamma  \right)  =   \frac{\alpha }{3}-\sqrt[3]{  -\frac{\alpha ^{3}}{27}+\frac{\beta \alpha }{6}-\frac{\gamma }{2}  +\sqrt{  \frac{\alpha ^{3}\gamma }{27} - \frac{\alpha ^{2}\beta ^{2}}{108} - \frac{\alpha \beta \gamma }{6}+\frac{\beta ^{3}}{27}+\frac{\gamma ^{2}}{4}}} -\sqrt[3]{  -\frac{\alpha ^{3}}{27}+\frac{\beta \alpha }{6}-\frac{\gamma }{2}  -\sqrt{\frac{\alpha ^{3}\gamma }{27}-\frac{\alpha ^{2}\beta ^{2}}{108 } - \frac{\alpha \beta \gamma }{6} + \frac{\beta ^{3}}{27} + \frac{\gamma ^{2}}{4}}  } $, $ \iota_{k^{\prime}}^{\left( n\right) } \left( \alpha ,\beta, \gamma, \lambda _{k}^{\left( n\right) }  \right) = -  \frac{\lambda _{k}^{\left( n\right) }-\alpha}{2}   + \sqrt{\left( \frac{\lambda _{k}^{\left( n\right) }-\alpha}{2} \right) ^{2}-\frac{\gamma }{\lambda _{k}^{\left( n\right) }}} $, and  $\varsigma _{k^{\prime\prime}}^{\left( n\right) } \left( \alpha ,\beta, \gamma, \lambda _{k}^{\left( n\right) }  \right) = -  \frac{\lambda _{k}^{\left( n\right) }-\alpha}{2}   - \sqrt{\left( \frac{\lambda _{k}^{\left( n\right) }-\alpha}{2} \right) ^{2}-\frac{\gamma }{\lambda _{k}^{\left( n\right) }}}$. Each eigenvalue displayed in this table corresponds with the  relaxation rates as denoted by  $R_{k}^{\left( n\right) } = -C  \lambda_{k}^{\left( n\right) } $.}
\label{tab:Eigenvalues}
\end{table}

\textit{Fourth coherence order:} the density matrix elements of $4$th order  denoted as  $\rho _{5,1}\left( t\right) $,  $\rho _{6,2}\left( t\right) $,  $\rho _{7,3}\left( t\right) $ and  $\rho _{8,4}\left( t\right) $ can be represented with the Eq. (\ref{RedfieldElementoGeral}) and (\ref{RedfieldElementoGeralInitialConditions}) such that the subscript indices $p$, $n$ and $q$ obey the values $p \equiv \left\{ 1 , 2 , 3 ,4\right\}$,  $n \equiv \left\{ 1 , 2 , 3 ,4 \right\}$ and the coherence order $q=4$. With this notation in hands, the transformation matrix $\mathbf{W}^{\left( 4\right)}$ must be resumed in matrix notation as denoted explicitly by
\begin{equation*}
\mathbf{W}^{\left( 4\right)}=\left[ 
\begin{array}{cccc}
W_{1,1}^{\left( 4\right)} & W_{1,2}^{\left( 4\right)} & W_{1,3}^{\left( 4\right)} & W_{1,4}^{\left( 4\right)} \\ 
W_{2,1}^{\left( 4\right)} & W_{2,2}^{\left( 4\right)} & W_{2,3}^{\left( 4\right)} & W_{2,4}^{\left( 4\right)} \\ 
W_{3,1}^{\left( 4\right)} & W_{3,2}^{\left( 4\right)} & W_{3,3}^{\left( 4\right)} & W_{3,4}^{\left( 4\right)} \\ 
W_{4,1}^{\left( 4\right)} & W_{4,2}^{\left( 4\right)} & W_{4,3}^{\left( 4\right)} & W_{4,4}^{\left( 4\right)}
\end{array}
\right] =\frac{1}{\sqrt{2}}\left[ 
\begin{array}{cccc}
\frac{\sqrt{a_{1}^{2}+1}}{a_{1}-b_{1}} & -\frac{b_{1}\sqrt{a_{1}^{2}+1}}{%
a_{1}-b_{1}} & -\frac{b_{1}\sqrt{a_{1}^{2}+1}}{a_{1}-b_{1}} & \frac{\sqrt{%
a_{1}^{2}+1}}{a_{1}-b_{1}} \\ 
-\frac{\sqrt{b_{1}^{2}+1}}{a_{1}-b_{1}} & \frac{a_{1}\sqrt{b_{1}^{2}+1}}{%
a_{1}-b_{1}} & \frac{a_{1}\sqrt{b_{1}^{2}+1}}{a_{1}-b_{1}} & -\frac{\sqrt{%
b_{1}^{2}+1}}{a_{1}-b_{1}} \\ 
\frac{b_{2}\sqrt{a_{2}^{2}+1}}{a_{2}-b_{2}} & -\frac{\sqrt{a_{2}^{2}+1}}{%
a_{2}-b_{2}} & \frac{\sqrt{a_{2}^{2}+1}}{a_{2}-b_{2}} & -\frac{b_{2}\sqrt{%
a_{2}^{2}+1}}{a_{2}-b_{2}} \\ 
-\frac{a_{2}\sqrt{b_{2}^{2}+1}}{a_{2}-b_{2}} & \frac{\sqrt{b_{2}^{2}+1}}{%
a_{2}-b_{2}} & -\frac{\sqrt{b_{2}^{2}+1}}{a_{2}-b_{2}} & \frac{a_{2}\sqrt{%
b_{2}^{2}+1}}{a_{2}-b_{2}}%
\end{array}%
\right] \text{,}
\end{equation*}
and the inverse transformation  $\overline{\mathbf{W}}^{ \left( 4\right)}$ denoted by
\begin{equation*}
\overline{\mathbf{W}}^{ \left( 4\right)}=\left[ 
\begin{array}{cccc}
\overline{W}_{1,1}^{ \left( 4\right) } & \overline{W}_{1,2}^{ \left( 4\right)} & \overline{W}_{1,3}^{ \left( 4\right) } & \overline{W}_{1,4}^{\left( 4\right) }
\\ 
\overline{W}_{2,1}^{ \left( 4\right)} & \overline{W}_{2,2}^{\left( 4\right) } & \overline{W}_{2,3}^{ \left( 4\right)} & \overline{W}_{2,4}^{ \left( 4\right) }
\\ 
\overline{W}_{3,1}^{\left( 4\right) } & \overline{W}_{3,2}^{ \left( 4\right) } & \overline{W}_{3,3}^{ \left( 4\right) } & \overline{W}_{3,4}^{ \left( 4\right) }
\\ 
\overline{W}_{4,1}^{ \left( 4\right)} & \overline{W}_{4,2}^{ \left( 4\right) } & \overline{W}_{4,3}^{ \left( 4\right) } & \overline{W}_{4,4}^{ \left( 4\right) }%
\end{array}%
\right] =\frac{1}{\sqrt{2}}\left[ 
\begin{array}{cccc}
\frac{a_{1}}{\sqrt{a_{1}^{2}+1}} & \frac{b_{1}}{\sqrt{b_{1}^{2}+1}} & -\frac{%
1}{\sqrt{a_{2}^{2}+1}} & -\frac{1}{\sqrt{b_{2}^{2}+1}} \\ 
\frac{1}{\sqrt{a_{1}^{2}+1}} & \frac{1}{\sqrt{b_{1}^{2}+1}} & -\frac{a_{2}}{%
\sqrt{a_{2}^{2}+1}} & -\frac{b_{2}}{\sqrt{b_{2}^{2}+1}} \\ 
\frac{1}{\sqrt{a_{1}^{2}+1}} & \frac{1}{\sqrt{b_{1}^{2}+1}} & \frac{a_{2}}{%
\sqrt{a_{2}^{2}+1}} & \frac{b_{2}}{\sqrt{b_{2}^{2}+1}} \\ 
\frac{a_{1}}{\sqrt{a_{1}^{2}+1}} & \frac{b_{1}}{\sqrt{b_{1}^{2}+1}} & \frac{1%
}{\sqrt{a_{2}^{2}+1}} & \frac{1}{\sqrt{b_{2}^{2}+1}}%
\end{array}%
\right] \text{,}
\end{equation*}
where the $a_{1}$, $b_{1}$, $a_{2}$, and $b_{2}$  coefficients are denoted by
\begin{eqnarray*}
a_{1} &=&\frac{16J_{0}-16J_{1}+\sqrt{%
256(J_{0}-J_{1})^{2}+105(J_{1}-J_{2})^{2}}}{\sqrt{%
105}\left( J_{1}-J_{2}\right) }\text{,} \\
b_{1} &=&\frac{16J_{0}-16J_{1}-\sqrt{%
256(J_{0}-J_{1})^{2}+105(J_{1}-J_{2})^{2}}}{\sqrt{%
105}\left( J_{1}-J_{2}\right) }\text{,}\\
a_{2} &=&\frac{-16J_{0}+\sqrt{256J_{0}^{2}+105\left( J_{1}+J_{2}\right) ^{2}}%
}{\sqrt{105}\left( J_{1}+J_{2}\right) } \\
b_{2} &=&\frac{-16J_{0}-\sqrt{256J_{0}^{2}+105\left( J_{1}+J_{2}\right) ^{2}}%
}{\sqrt{105}\left( J_{1}+J_{2}\right) }
\end{eqnarray*}
therefore, applying the general notation of Eq. (\ref{RedfieldElementoGeral}) and considering the inital time $t_{0}=0$, the fourth order density matrix elements are denoted by
\begin{eqnarray*}
\rho _{5,1}\left( t\right)  &=&\overline{W}_{1,1}^{ \left( 4\right) }\exp \left[ -R_{1}^{\left(
4\right) } \ t \   \right] \widetilde{\rho }_{1}\left(
0\right) +\overline{W}_{1,2}^{ \left( 4\right) }\exp \left[ -R_{2}^{\left( 4\right) }
 \ t \   \right] \widetilde{\rho }_{2}\left( 0\right)
+\overline{W}_{1,3}^{ \left( 4\right)}\exp \left[ -R_{3}^{\left( 4\right) } \ t \  
\right] \widetilde{\rho }_{3}\left( 0\right) +\overline{W}_{1,4}^{ \left( 4\right) }\exp %
\left[ -R_{4}^{\left( 4\right) } \ t \   \right] \widetilde{%
\rho }_{4}\left( 0\right) \text{,} \\
\rho _{6,2}\left( t\right)  &=&\overline{W}_{2,1}^{ \left( 4\right) }\exp \left[ -R_{1}^{\left(
4\right) } \ t \   \right] \widetilde{\rho }_{1}\left(
0\right) +\overline{W}_{2,2}^{ \left( 4\right) }\exp \left[ -R_{2}^{\left( 4\right) } 
 \ t \   \right] \widetilde{\rho }_{2}\left( 0\right)
+\overline{W}_{2,3}^{ \left( 4\right)}\exp \left[ -R_{3}^{\left( 4\right) } \ t \   
\right] \widetilde{\rho }_{3}\left( 0\right) +\overline{W}_{2,4}^{ \left( 4\right) }\exp %
\left[ -R_{4}^{\left( 4\right) } \ t \   \right] \widetilde{%
\rho }_{4}\left( 0\right) \text{,} \\
\rho _{7,3}\left( t\right)  &=&\overline{W}_{3,1}^{ \left( 4\right) }\exp \left[ -R_{1}^{\left(
4\right) } \ t \   \right] \widetilde{\rho }_{1}\left(
0\right) +\overline{W}_{3,2}^{ \left( 4\right)}\exp \left[ -R_{2}^{\left( 4\right) } 
 \ t \   \right] \widetilde{\rho }_{2}\left( 0\right)
+\overline{W}_{3,3}^{ \left( 4\right)}\exp \left[ -R_{3}^{\left( 4\right) } \ t \   %
\right] \widetilde{\rho }_{3}\left( 0\right) +\overline{W}_{3,4}^{ \left( 4\right) }\exp %
\left[ -R_{4}^{\left( 4\right) } \ t \   \right] \widetilde{%
\rho }_{4}\left( 0\right) \text{,} \\
\rho _{8,4}\left( t\right)  &=&\overline{W}_{4,1}^{ \left( 4\right) }\exp \left[ -R_{1}^{\left(
4\right) } \ t \   \right] \widetilde{\rho }_{1}\left(
0\right) +\overline{W}_{4,2}^{ \left( 4\right) }\exp \left[ -R_{2}^{\left( 4\right) } 
 \ t \   \right] \widetilde{\rho }_{2}\left( 0\right)
+\overline{W}_{4,3}^{ \left( 4\right) }\exp \left[ -R_{3}^{\left( 4\right) } \ t \   %
\right] \widetilde{\rho }_{3}\left( 0\right) +\overline{W}_{4,4}^{ \left( 4\right) }\exp %
\left[ -R_{4}^{\left( 4\right) } \ t \   \right] \widetilde{%
\rho }_{4}\left( 0\right) \text{,}
\end{eqnarray*}
where the relaxation rates of $4$th order are resumed at their eigenvalue notation in Tab.  \ref{tab:Eigenvalues}. Also, the relaxation rates $R_{3}^{\left( 4\right) }$ and $R_{4}^{\left( 4\right) }$   are analytical and compatible solutions of the relaxation operator denoted at  Eq. (A6a) of Ref. \cite{tsoref1996}. The relaxation rates  $R_{1}^{\left( 4\right) }$ and $R_{2}^{\left( 4\right) }$  are the new ones introduced to describe the dynamics of $4$th coherence order density matrix elements.

\textit{Third coherences order:}  the density matrix elements of $3$th order  denoted as $\rho _{4,1}\left( t\right)$,  $\rho _{5,2}\left( t\right)$,  $\rho _{6,3}\left( t\right)$, $\rho _{7,4}\left( t\right)$, and  $\rho _{8,5}\left( t\right)$  can be represented with the Eq. (\ref{RedfieldElementoGeral}) and (\ref{RedfieldElementoGeralInitialConditions}) such that the subscript indices  $p$, $n$ and $q$ obey the values    $p \equiv \left\{ 1 , 2 , 3 , 4 , 5\right\}$,  $n \equiv \left\{ 1 , 2 , 3 , 4 , 5 \right\}$ and the coherence order $q=3$. With this notation in hands, the transformation  $\mathbf{W}^{\left( 3\right)}$ must be resumed in matrix notation as denoted explicitly by
\begin{eqnarray*}
\mathbf{W}^{\left( 3\right)}&=&\left[ 
\begin{array}{cc}
\frac{\sqrt{91}\left( z_{5}y_{3}-y_{5}z_{3}\right) +\sqrt{30}\left(
y_{5}x_{3}-x_{5}y_{3}\right) +2\sqrt{77}\left( x_{5}z_{3}-z_{5}x_{3}\right) 
}{\sqrt{858}\left(
x_{5}y_{1}z_{3}-x_{5}y_{3}z_{1}-y_{5}x_{1}z_{3}+y_{5}x_{3}z_{1}+z_{5}x_{1}y_{3}-z_{5}x_{3}y_{1}\right) 
} & \frac{-4\sqrt{13}\left( y_{5}z_{3}-z_{5}y_{3}\right) -\sqrt{11}\left(
x_{5}z_{3}-z_{5}x_{3}\right) +\sqrt{210}\left( x_{5}y_{3}-y_{5}x_{3}\right) 
}{\sqrt{858}\left(
x_{5}y_{1}z_{3}-x_{5}y_{3}z_{1}-y_{5}x_{1}z_{3}+y_{5}x_{3}z_{1}+z_{5}x_{1}y_{3}-z_{5}x_{3}y_{1}\right) 
} \\ 
0 & -\frac{\sqrt{3}\sqrt{11}\sqrt{13}}{\sqrt{858}} \\ 
\frac{\sqrt{91}\left( y_{5}z_{1}-z_{5}y_{1}\right) +\sqrt{30}\left(
x_{5}y_{1}-y_{5}x_{1}\right) +2\sqrt{77}\left( z_{5}x_{1}-x_{5}z_{1}\right) 
}{\sqrt{858}\left(
x_{5}y_{1}z_{3}-x_{5}y_{3}z_{1}-y_{5}x_{1}z_{3}+y_{5}x_{3}z_{1}+z_{5}x_{1}y_{3}-z_{5}x_{3}y_{1}\right) 
} & \frac{\sqrt{11}\left( x_{5}z_{1}-z_{5}x_{1}\right) +4\sqrt{13}\left(
y_{5}z_{1}-z_{5}y_{1}\right) -\sqrt{210}\left( x_{5}y_{1}-y_{5}x_{1}\right) 
}{\sqrt{858}\left(
x_{5}y_{1}z_{3}-x_{5}y_{3}z_{1}-y_{5}x_{1}z_{3}+y_{5}x_{3}z_{1}+z_{5}x_{1}y_{3}-z_{5}x_{3}y_{1}\right) 
} \\ 
-\frac{\sqrt{13}\sqrt{3}\sqrt{11}}{\sqrt{858}} & 0 \\ 
\frac{\sqrt{91}\left( y_{1}z_{3}-y_{3}z_{1}\right) +\sqrt{30}\left(
x_{1}y_{3}-x_{3}y_{1}\right) +2\sqrt{77}\left( x_{3}z_{1}-x_{1}z_{3}\right) 
}{\sqrt{858}\left(
x_{5}y_{1}z_{3}-x_{5}y_{3}z_{1}-y_{5}x_{1}z_{3}+y_{5}x_{3}z_{1}+z_{5}x_{1}y_{3}-z_{5}x_{3}y_{1}\right) 
} & \frac{\sqrt{11}\left( x_{1}z_{3}-x_{3}z_{1}\right) +4\sqrt{13}\left(
y_{1}z_{3}-y_{3}z_{1}\right) -\sqrt{210}\left( x_{1}y_{3}-x_{3}y_{1}\right) 
}{\sqrt{858}\left(
x_{5}y_{1}z_{3}-x_{5}y_{3}z_{1}-y_{5}x_{1}z_{3}+y_{5}x_{3}z_{1}+z_{5}x_{1}y_{3}-z_{5}x_{3}y_{1}\right) 
}%
\end{array}%
\right. \cdots  \\
&&\left. 
\begin{array}{cc}
\frac{-3\sqrt{42}\left( x_{5}y_{3}-y_{5}x_{3}\right) -2\sqrt{65}\left(
y_{5}z_{3}-z_{5}y_{3}\right) -2\sqrt{55}\left( x_{5}z_{3}-z_{5}x_{3}\right) 
}{\sqrt{858}\left(
x_{5}y_{1}z_{3}-x_{5}y_{3}z_{1}-y_{5}x_{1}z_{3}+y_{5}x_{3}z_{1}+z_{5}x_{1}y_{3}-z_{5}x_{3}y_{1}\right) 
} & \frac{\sqrt{210}\left( x_{5}y_{3}-y_{5}x_{3}\right) -4\sqrt{13}\left(
y_{5}z_{3}-z_{5}y_{3}\right) -\sqrt{11}\left( x_{5}z_{3}-z_{5}x_{3}\right) }{%
\sqrt{858}\left(
x_{5}y_{1}z_{3}-x_{5}y_{3}z_{1}-y_{5}x_{1}z_{3}+y_{5}x_{3}z_{1}+z_{5}x_{1}y_{3}-z_{5}x_{3}y_{1}\right) 
} \\ 
0 & \frac{\sqrt{3}\sqrt{11}\sqrt{13}}{\sqrt{858}} \\ 
\frac{3\sqrt{42}\left( x_{5}y_{1}-y_{5}x_{1}\right) +2\sqrt{65}\left(
y_{5}z_{1}-z_{5}y_{1}\right) +2\sqrt{55}\left( x_{5}z_{1}-z_{5}x_{1}\right) 
}{\sqrt{858}\left(
x_{5}y_{1}z_{3}-x_{5}y_{3}z_{1}-y_{5}x_{1}z_{3}+y_{5}x_{3}z_{1}+z_{5}x_{1}y_{3}-z_{5}x_{3}y_{1}\right) 
} & \frac{\sqrt{11}\left( x_{5}z_{1}-z_{5}x_{1}\right) +4\sqrt{13}\left(
y_{5}z_{1}-z_{5}y_{1}\right) -\sqrt{210}\left( x_{5}y_{1}-y_{5}x_{1}\right) 
}{\sqrt{858}\left(
x_{5}y_{1}z_{3}-x_{5}y_{3}z_{1}-y_{5}x_{1}z_{3}+y_{5}x_{3}z_{1}+z_{5}x_{1}y_{3}-z_{5}x_{3}y_{1}\right) 
} \\ 
0 & 0 \\ 
\frac{3\sqrt{42}\left( x_{1}y_{3}-x_{3}y_{1}\right) +2\sqrt{65}\left(
y_{1}z_{3}-y_{3}z_{1}\right) +2\sqrt{55}\left( x_{1}z_{3}-x_{3}z_{1}\right) 
}{\sqrt{858}\left(
x_{5}y_{1}z_{3}-x_{5}y_{3}z_{1}-y_{5}x_{1}z_{3}+y_{5}x_{3}z_{1}+z_{5}x_{1}y_{3}-z_{5}x_{3}y_{1}\right) 
} & \frac{\sqrt{11}\left( x_{1}z_{3}-x_{3}z_{1}\right) +4\sqrt{13}\left(
y_{1}z_{3}-y_{3}z_{1}\right) -\sqrt{210}\left( x_{1}y_{3}-x_{3}y_{1}\right) 
}{\sqrt{858}\left(
x_{5}y_{1}z_{3}-x_{5}y_{3}z_{1}-y_{5}x_{1}z_{3}+y_{5}x_{3}z_{1}+z_{5}x_{1}y_{3}-z_{5}x_{3}y_{1}\right) 
}%
\end{array}%
\right. \cdots  \\
&&\left. 
\begin{array}{c}
\frac{-\sqrt{30}\left( x_{5}y_{3}-y_{5}x_{3}\right) -\sqrt{91}\left(
y_{5}z_{3}-z_{5}y_{3}\right) +2\sqrt{77}\left( x_{5}z_{3}-z_{5}x_{3}\right) 
}{\sqrt{858}\left(
x_{5}y_{1}z_{3}-x_{5}y_{3}z_{1}-y_{5}x_{1}z_{3}+y_{5}x_{3}z_{1}+z_{5}x_{1}y_{3}-z_{5}x_{3}y_{1}\right) 
} \\ 
0 \\ 
\frac{\sqrt{91}\left( y_{5}z_{1}-z_{5}y_{1}\right) +\sqrt{30}\left(
x_{5}y_{1}-y_{5}x_{1}\right) -2\sqrt{77}\left( x_{5}z_{1}-z_{5}x_{1}\right) 
}{\sqrt{858}\left(
x_{5}y_{1}z_{3}-x_{5}y_{3}z_{1}-y_{5}x_{1}z_{3}+y_{5}x_{3}z_{1}+z_{5}x_{1}y_{3}-z_{5}x_{3}y_{1}\right) 
} \\ 
\frac{\sqrt{3}\sqrt{11}\sqrt{13}}{\sqrt{858}} \\ 
\frac{\sqrt{91}\left( y_{1}z_{3}-y_{3}z_{1}\right) +\sqrt{30}\left(
x_{1}y_{3}-x_{3}y_{1}\right) -2\sqrt{77}\left( x_{1}z_{3}-x_{3}z_{1}\right) 
}{\sqrt{858}\left(
x_{5}y_{1}z_{3}-x_{5}y_{3}z_{1}-y_{5}x_{1}z_{3}+y_{5}x_{3}z_{1}+z_{5}x_{1}y_{3}-z_{5}x_{3}y_{1}\right) 
}%
\end{array}%
\right] \text{,}
\end{eqnarray*}
and the inverse transformation $\overline{\mathbf{W}}^{ \left( 3\right)}$  denoted by
\begin{equation*}
\overline{\mathbf{W}}^{ \left( 3\right)} = \left[ 
\begin{array}{ccccc}
\frac{\sqrt{462}}{66}x_{1}+\frac{\sqrt{546}}{39}y_{1}+\frac{\sqrt{715}}{143}%
z_{1} & 0 & \frac{\sqrt{462}}{66}x_{3}+\frac{\sqrt{546}}{39}y_{3}+\frac{%
\sqrt{715}}{143}z_{3} & -\frac{1}{\sqrt{2}} & \frac{\sqrt{462}}{66}x_{5}+%
\frac{\sqrt{546}}{39}y_{5}+\frac{\sqrt{715}}{143}z_{5} \\ 
\frac{\sqrt{264}}{33}x_{1}-\frac{\sqrt{78}}{78}y_{1}-\frac{\sqrt{5005}}{143}%
z_{1} & -\frac{1}{\sqrt{2}} & \frac{\sqrt{264}}{33}x_{3}-\frac{\sqrt{78}}{78}%
y_{3}-\frac{\sqrt{5005}}{143}z_{3} & 0 & \frac{\sqrt{264}}{33}x_{5}-\frac{%
\sqrt{78}}{78}y_{5}-\frac{\sqrt{5005}}{143}z_{5} \\ 
\frac{\sqrt{330}}{33}x_{1}-\frac{\sqrt{390}}{39}y_{1}+\frac{\sqrt{9009}}{143}%
z_{1} & 0 & \frac{\sqrt{330}}{33}x_{3}-\frac{\sqrt{390}}{39}y_{3}+\frac{%
\sqrt{9009}}{143}z_{3} & 0 & \frac{\sqrt{330}}{33}x_{5}-\frac{\sqrt{390}}{39}%
y_{5}+\frac{\sqrt{9009}}{143}z_{5} \\ 
\frac{\sqrt{264}}{33}x_{1}-\frac{\sqrt{78}}{78}y_{1}-\frac{\sqrt{5005}}{143}%
z_{1} & \frac{1}{\sqrt{2}} & \frac{\sqrt{264}}{33}x_{3}-\frac{\sqrt{78}}{78}%
y_{3}-\frac{\sqrt{5005}}{143}z_{3} & 0 & \frac{\sqrt{264}}{33}x_{5}-\frac{%
\sqrt{78}}{78}y_{5}-\frac{\sqrt{5005}}{143}z_{5} \\ 
\frac{\sqrt{462}}{66}x_{1}+\frac{\sqrt{546}}{39}y_{1}+\frac{\sqrt{715}}{143}%
z_{1} & 0 & \frac{\sqrt{462}}{66}x_{3}+\frac{\sqrt{546}}{39}y_{3}+\frac{%
\sqrt{715}}{143}z_{3} & \frac{1}{\sqrt{2}} & \frac{\sqrt{462}}{66}x_{5}+%
\frac{\sqrt{546}}{39}y_{5}+\frac{\sqrt{715}}{143}z_{5}%
\end{array}%
\right] \text{,}
\end{equation*}
where the cartesian parameters are denoted by
\begin{equation*}
\mathbf{r}_{k}=\left[ 
\begin{array}{c}
x_{k} \\ 
y_{k} \\ 
z_{k}%
\end{array}%
\right] = \frac{1}{\sqrt{\xi _{4}^{2}\left( \lambda _{k}^{\left( 3\right) }-\xi _{3}\right) ^{2}+\left( \lambda _{k}^{\left( 3\right) }-\xi _{1}\right) ^{2}\left( \lambda _{k}^{\left( 3\right) }-\xi _{3}\right) ^{2}+\xi _{5}^{2}\left( \lambda _{k}^{\left( 3\right) }-\xi _{1}\right) ^{2}}}\left[ 
\begin{array}{c}
\xi _{4}\left( \lambda _{k}^{\left( 3\right) }-\xi _{3}\right) \\ 
\left( \lambda _{k}^{\left( 3\right) }-\xi _{1}\right) \left( \lambda _{k}^{\left( 3\right) }-\xi _{3}\right) \\ 
\xi _{5}\left( \lambda _{k}^{\left( 3\right) }-\xi _{1}\right)%
\end{array}%
\right]\text{,} 
\end{equation*}
with the subscript indices are $ k=1,3,5 $ and the $\xi_{s}$-parameters are resumed by
\begin{eqnarray*}
\xi _{1} &=& -12J_{0}-29J_{1}-9J_{2} \\
\xi _{2} &=& \frac{1}{13}\left( -339J_{0}-225J_{1}-476J_{2}\right)  \text{,}  \\
\xi _{3} &=& \frac{1}{13}\left( -90J_{0}-113J_{1}-161J_{2}\right)  \text{,}  \\
\xi _{4} &=&  \sqrt{\frac{208}{11}}\left( -3J_{0}+2J_{1}+J_{2}\right)  \text{,}  \\
\xi _{5} &=& - \sqrt{\frac{210}{1859}}\left( 27J_{0}+4J_{1}-31J_{2}\right)   \text{,} 
\end{eqnarray*}
therefore, applying the general notation of Eq. (\ref{RedfieldElementoGeral}) and considering the initial time $t_{0}=0$, the third order density matrix elements are denoted by
\begin{eqnarray*}
\rho _{4,1}\left( t\right)  &=&\overline{W}_{1,1}^{ \left( 3\right)}\exp \left[ -R_{1}^{\left(
3\right) } \ t \ \right] \widetilde{\rho }_{1}\left(
0\right) +\overline{W}_{1,3}^{ \left( 3\right) }\exp \left[ -R_{3}^{\left( 3\right) } \
t \ \right] \widetilde{\rho }_{3}\left( 0\right)
+\overline{W}_{1,4}^{  \left( 3\right)}\exp \left[ -R_{4}^{\left( 3\right) } \ t \  %
\right] \widetilde{\rho }_{4}\left( 0\right) +\overline{W}_{1,5}^{  \left( 3\right) }\exp %
\left[ -R_{5}^{\left( 3\right) } \ t \  \right] \widetilde{%
\rho }_{5}\left( 0\right) \text{,} \\
\rho _{5,2}\left( t\right)  &=&\overline{W}_{2,1}^{  \left( 3\right) }\exp \left[ -R_{1}^{\left(
3\right) } \ t \  \right] \widetilde{\rho }_{1}\left(
0\right) +\overline{W}_{2,2}^{  \left( 3\right) }\exp \left[ -R_{2}^{\left( 3\right) } \ t \  \right] \widetilde{\rho }_{2}\left( 0\right)
+\overline{W}_{2,3}^{  \left( 3\right) }\exp \left[ -R_{3}^{\left( 3\right) } \ t \  %
\right] \widetilde{\rho }_{3}\left( 0\right) +\overline{W}_{2,5}^{  \left( 3\right) }\exp %
\left[ -R_{5}^{\left( 3\right) } \ t \  \right] \widetilde{%
\rho }_{5}\left( 0\right) \text{,} \\
\rho _{6,3}\left( t\right)  &=&\overline{W}_{3,1}^{\left( 3\right) }\exp \left[ -R_{1}^{\left(
3\right) } \ t \  \right] \widetilde{\rho }_{1}\left(
0\right) +\overline{W}_{3,3}^{  \left( 3\right)}\exp \left[ -R_{3}^{\left( 3\right) }
 \ t \  \right] \widetilde{\rho }_{3}\left( 0\right)
+\overline{W}_{3,5}^{  \left( 3\right) }\exp \left[ -R_{5}^{\left( 3\right) } \ t \  %
\right] \widetilde{\rho }_{5}\left(0\right) \text{,} \\
\rho _{7,4}\left( t\right)  &=&\overline{W}_{4,1}^{  \left( 3\right) }\exp \left[ -R_{1}^{\left(
3\right) } \ t \  \right] \widetilde{\rho }_{1}\left(
0\right) +\overline{W}_{4,2}^{  \left( 3\right)}\exp \left[ -R_{2}^{\left( 3\right) } 
 \ t \  \right] \widetilde{\rho }_{2}\left( 0\right)
+\overline{W}_{4,3}^{ \left( 3\right)}\exp \left[ -R_{3}^{\left( 3\right) } \ t \  %
\right] \widetilde{\rho }_{3}\left( 0\right) +\overline{W}_{4,5}^{  \left( 3\right) }\exp %
\left[ -R_{5}^{\left( 3\right) } \ t \  \right] \widetilde{%
\rho }_{5}\left( 0\right) \text{,} \\
\rho _{8,5}\left( t\right)  &=&\overline{W}_{5,1}^{ \left( 3\right)}\exp \left[ -R_{1}^{\left(
3\right) } \ t \  \right] \widetilde{\rho }_{1}\left(
0\right) +\overline{W}_{5,3}^{ \left( 3\right)}\exp \left[ -R_{3}^{\left( 3\right) } 
 \ t \  \right] \widetilde{\rho }_{3}\left( 0\right)
+\overline{W}_{5,4}^{  \left( 3\right)}\exp \left[ -R_{4}^{\left( 3\right) } \ t \  %
\right] \widetilde{\rho }_{4}\left( 0\right) +\overline{W}_{5,5}^{  \left( 3\right) }\exp %
\left[ -R_{5}^{\left( 3\right) } \ t \  \right] \widetilde{%
\rho }_{5}\left( 0\right) \text{,}
\end{eqnarray*}
where the relaxation rates of $3$th order are resumed at their eigenvalue notation in Tab.  \ref{tab:Eigenvalues}. Also, the relaxation rates $R_{1}^{\left( 3\right) }$, $R_{3}^{\left( 3\right) }$ and $R_{5}^{\left( 3\right) }$   are analytical and compatible solutions of the relaxation operator denoted at  Eq. (A5a) of Ref. \cite{tsoref1996}. The relaxation rates  $R_{2}^{\left( 4\right) }$ and $R_{4}^{\left( 4\right) }$  are the new ones introduced to describe the dynamics of $3$th coherence order density matrix elements.

\textit{Second coherence order:} the density matrix elements of $2$th order  denoted as  $\rho _{3,1}\left( t\right)$,  $\rho _{4,2}\left( t\right)$,  $\rho _{5,3}\left( t\right)$, $\rho _{6,4}\left( t\right)$,    $\rho _{7,5}\left( t\right)$  and  $\rho _{8,6}\left( t\right)$  can be represented with the Eq. (\ref{RedfieldElementoGeral}) and (\ref{RedfieldElementoGeralInitialConditions}) such that the subscript indices $p$, $n$ and $q$ obey the values $p \equiv \left\{ 1 , 2 , 3 , 4 , 5 , 6 \right\}$,  $n \equiv \left\{ 1 , 2 , 3 , 4 , 5 , 6 \right\}$ and the  coherence order $q=2$. With this notation in hands, the transformation  $\mathbf{W} ^{\left( 2\right)}$ must be resumed in matrix notation as denoted by

\begin{small}
\begin{equation*}
\mathbf{W}^{\left( 2\right)} =  \left[ 
\begin{array}{cccccc}
\frac{z_{5}y_{2}-y_{5}z_{2}}{M_{1,2,5}} & \frac{x_{5}z_{2}-z_{5}x_{2}}{M_{1,2,5}}
& 0 & 0 & \frac{y_{5}x_{2}-x_{5}y_{2}}{M_{1,2,5}} & 0 \\ 
\frac{y_{5}z_{1}-z_{5}y_{1}}{M_{1,2,5}} & \frac{z_{5}x_{1}-x_{5}z_{1}}{M_{1,2,5}}
& 0 & 0 & \frac{x_{5}y_{1}-y_{5}x_{1}}{M_{1,2,5}} & 0 \\ 
0 & 0 & \frac{y_{4}z_{6}-y_{6}z_{4}}{M_{3,4,6}} & \frac{x_{6}z_{4}-x_{4}z_{6}}{%
M_{3,4,6}} & 0 & \frac{x_{4}y_{6}-x_{6}y_{4}}{M_{3,4,6}} \\ 
0 & 0 & \frac{z_{3}y_{6}-y_{3}z_{6}}{M_{3,4,6}} & \frac{x_{3}z_{6}-x_{6}z_{3}}{%
M_{3,4,6}} & 0 & \frac{y_{3}x_{6}-x_{3}y_{6}}{M_{3,4,6}} \\ 
\frac{y_{1}z_{2}-y_{2}z_{1}}{M_{1,2,5}} & \frac{x_{2}z_{1}-x_{1}z_{2}}{M_{1,2,5}}
& 0 & 0 & \frac{x_{1}y_{2}-x_{2}y_{1}}{M_{1,2,5}} & 0 \\ 
0 & 0 & \frac{y_{3}z_{4}-y_{4}z_{3}}{M_{3,4,6}} & \frac{x_{4}z_{3}-x_{3}z_{4}}{%
M_{3,4,6}} & 0 & \frac{x_{3}y_{4}-x_{4}y_{3}}{M_{3,4,6}}%
\end{array}%
\right] \left[ 
\begin{array}{cccccc}
\frac{\sqrt{5}}{\sqrt{66}} & -\frac{\sqrt{21}}{\sqrt{66}} & \frac{\sqrt{7}}{%
\sqrt{66}} & \frac{\sqrt{7}}{\sqrt{66}} & -\frac{\sqrt{21}}{\sqrt{66}} & 
\frac{\sqrt{5}}{\sqrt{66}} \\ 
\frac{\sqrt{3}}{6} & \frac{\sqrt{35}}{14} & \frac{\sqrt{105}}{21} & \frac{%
\sqrt{105}}{21} & \frac{\sqrt{35}}{14} & \frac{\sqrt{3}}{6} \\ 
-\frac{\sqrt{1155}}{66} & -\frac{\sqrt{99}}{22} & -\frac{\sqrt{33}}{33} & 
\frac{\sqrt{33}}{33} & \frac{\sqrt{99}}{22} & \frac{\sqrt{1155}}{66} \\ 
-\frac{\sqrt{3}}{\sqrt{286}} & \frac{\sqrt{35}}{\sqrt{286}} & -\frac{\sqrt{%
105}}{\sqrt{286}} & \frac{\sqrt{105}}{\sqrt{286}} & -\frac{\sqrt{35}}{\sqrt{%
286}} & \frac{\sqrt{3}}{\sqrt{286}} \\ 
\frac{\sqrt{15}}{2\sqrt{11}} & \frac{\sqrt{77}}{154} & -\frac{2\sqrt{231}}{77%
} & -\frac{2\sqrt{231}}{77} & \frac{\sqrt{77}}{154} & \frac{\sqrt{15}}{2%
\sqrt{11}} \\ 
-\frac{\sqrt{35}}{2\sqrt{39}} & \frac{\sqrt{117}}{26} & \frac{2}{\sqrt{39}}
& -\frac{2}{\sqrt{39}} & -\frac{\sqrt{117}}{26} & \frac{\sqrt{1365}}{78}%
\end{array}%
\right] \text{,}
\end{equation*}
\end{small}

\begin{eqnarray*}
M_{1,2,5}
&=&x_{5}y_{1}z_{2}-x_{5}y_{2}z_{1}-y_{5}x_{1}z_{2}+y_{5}x_{2}z_{1}+z_{5}x_{1}y_{2}-z_{5}x_{2}y_{1}%
\text{,} \\
M_{3,4,6}
&=&x_{3}y_{4}z_{6}-x_{3}y_{6}z_{4}-x_{4}y_{3}z_{6}+x_{4}z_{3}y_{6}+y_{3}x_{6}z_{4}-x_{6}y_{4}z_{3}
\end{eqnarray*}
and the inverse transformation  denoted by $\overline{\mathbf{W}}^{\left( 2\right)}$ is
\begin{equation*}
\overline{\mathbf{W}}^{\left( 2\right)}  =  \left[ 
\begin{array}{cccccc}
\frac{\sqrt{5}}{\sqrt{66}} & \frac{\sqrt{7}}{\sqrt{7}\sqrt{12}} & -\frac{%
\sqrt{5}\sqrt{7}}{\sqrt{132}} & -\frac{\sqrt{3}}{\sqrt{286}} & \frac{\sqrt{7}%
\sqrt{3}\sqrt{5}}{\sqrt{7}\sqrt{44}} & -\frac{\sqrt{5}\sqrt{7}}{\sqrt{156}}
\\ 
-\frac{\sqrt{3}\sqrt{7}}{\sqrt{66}} & \frac{\sqrt{3}\sqrt{5}}{\sqrt{7}\sqrt{%
12}} & -\frac{3\sqrt{3}}{\sqrt{132}} & \frac{\sqrt{5}\sqrt{7}}{\sqrt{286}} & 
\frac{1}{\sqrt{7}\sqrt{44}} & \frac{3\sqrt{3}}{\sqrt{156}} \\ 
\frac{\sqrt{7}}{\sqrt{66}} & \frac{2\sqrt{5}}{\sqrt{7}\sqrt{12}} & -\frac{2}{%
\sqrt{132}} & -\frac{\sqrt{5}\sqrt{7}\sqrt{3}}{\sqrt{286}} & -\frac{4\sqrt{3}%
}{\sqrt{7}\sqrt{44}} & \frac{4}{\sqrt{156}} \\ 
\frac{\sqrt{7}}{\sqrt{66}} & \frac{2\sqrt{5}}{\sqrt{7}\sqrt{12}} & \frac{2}{%
\sqrt{132}} & \frac{\sqrt{5}\sqrt{7}\sqrt{3}}{\sqrt{286}} & -\frac{4\sqrt{3}%
}{\sqrt{7}\sqrt{44}} & -\frac{4}{\sqrt{156}} \\ 
-\frac{\sqrt{3}\sqrt{7}}{\sqrt{66}} & \frac{\sqrt{3}\sqrt{5}}{\sqrt{7}\sqrt{%
12}} & \frac{3\sqrt{3}}{\sqrt{132}} & -\frac{\sqrt{5}\sqrt{7}}{\sqrt{286}} & 
\frac{1}{\sqrt{7}\sqrt{44}} & -\frac{3\sqrt{3}}{\sqrt{156}} \\ 
\frac{\sqrt{5}}{\sqrt{66}} & \frac{\sqrt{7}}{\sqrt{7}\sqrt{12}} & \frac{%
\sqrt{5}\sqrt{7}}{\sqrt{132}} & \frac{\sqrt{3}}{\sqrt{286}} & \frac{\sqrt{7}%
\sqrt{3}\sqrt{5}}{\sqrt{7}\sqrt{44}} & \frac{\sqrt{5}\sqrt{7}}{\sqrt{156}}%
\end{array}%
\right] \left[ 
\begin{array}{cccccc}
x_{1} & x_{2} & 0 & 0 & x_{5} & 0 \\ 
y_{1} & y_{2} & 0 & 0 & y_{5} & 0 \\ 
0 & 0 & x_{3} & x_{4} & 0 & x_{6} \\ 
0 & 0 & y_{3} & y_{4} & 0 & y_{6} \\ 
z_{1} & z_{2} & 0 & 0 & z_{5} & 0 \\ 
0 & 0 & z_{3} & z_{4} & 0 & z_{6}%
\end{array}%
\right] \text{,}
\end{equation*}
where the cartesian parameters are denoted by
\begin{equation*}
\mathbf{r}=\left[ 
\begin{array}{c}
x_{k} \\ 
y_{k} \\ 
z_{k}
\end{array}%
\right] =\frac{1}{\sqrt{\xi _{4}^{2}\left( \lambda_{k}^{\left( 2\right) }-\xi
_{2}\right) ^{2}+\xi _{5}^{2}\left( \lambda_{k}^{\left( 2\right) }-\xi
_{1}\right) ^{2}+\left( \lambda_{k}^{\left( 2\right) }-\xi _{1}\right)
^{2}\left( \lambda_{k}^{\left( 2\right) }-\xi _{2}\right) ^{2}}}\left[ 
\begin{array}{c}
\xi _{4}\left( \lambda_{k}^{\left( 2\right) }-\xi _{2}\right)  \\ 
\xi _{5}\left( \lambda_{k}^{\left( 2\right) }-\xi _{1}\right)  \\ 
\left( \lambda_{k}^{\left( 2\right) }-\xi _{1}\right) \left( \lambda_{k}^{\left(
2\right) }-\xi _{2}\right) 
\end{array}%
\right] \text{,}
\end{equation*}
such that the computation of those components for the subscript index $k=1,2,5$ must be used the set of  $\xi_{s}$-parameters found at the left hand side of Eq. (\ref{CartesianParameters2ndOrder}), and for the subscript index   $k=3,4,6$ must be used the ones found at the right  hand side of Eq. (\ref{CartesianParameters2ndOrder})

\begin{equation}
\left\{ 
\begin{array}{ccl}
\xi _{1} & = &  -\frac{1}{11} \left(
107J_{0}+294J_{1}+369J_{2}\right) \text{,} \\ 
\xi _{2} & = & -\frac{1}{7} \left(
55J_{0}+102J_{1}+39J_{2}\right) \text{,} \\ 
\xi _{3} & = & - \frac{-1}{77} \left(
1341J_{0}+1440J_{1}+2609J_{2}\right) \text{,} \\ 
\xi _{4} & = &  -\frac{56}{11} \sqrt{2}\left(
J_{0}+J_{1}-2J_{2}\right) \text{,} \\ 
\xi _{5} & = & - \frac{4}{7} \sqrt{55}\left(
2J_{0}-J_{1}-J_{2}\right) \text{,}%
\end{array}%
\right. \qquad \qquad \qquad \left\{ 
\begin{array}{ccl}
\xi _{1} & = & - \frac{1}{3} \left(
51J_{0}+32J_{1}+67J_{2}\right) \text{,} \\ 
\xi _{2} & = & - \frac{1}{13} \left(
45J_{0}+168J_{1}+151J_{2}\right) \text{,} \\ 
\xi _{3} & = &- \frac{1}{39} \left(
567J_{0}+1030J_{1}+1523J_{2}\right) \text{,} \\ 
\xi _{4} & = & - \frac{4}{33} \sqrt{143}\left(
6J_{0}+J_{1}-7J_{2}\right) \text{,} \\ 
\xi _{5} & = & - \frac{24}{143} \sqrt{770}\left(
J_{0}+2J_{1}-3J_{2}\right) \text{,}%
\end{array}%
\right.   \label{CartesianParameters2ndOrder}
\end{equation}
therefore, applying the general notation of Eq. (\ref{RedfieldElementoGeral}), the second order density matrix elements are denoted by
\begin{equation}
\rho _{2+n,n}\left( t\right)   = \sum_{p=1}^{6} \overline{W}_{n,p}^{{\left( 2\right)} }\exp \left[ -R_{p}^{\left( 2\right) }\left( t-t_{0}\right) \right] \widetilde{\rho }_{p}\left( t_{0}\right)  \text{,}
\end{equation}
where the relaxation rates of $2$th order are resumed at their eigenvalue notation in Tab.  \ref{tab:Eigenvalues}. Also, the relaxation rates $R_{3}^{\left( 2\right) }$, $R_{4}^{\left( 2\right) }$ and $R_{6}^{\left( 2\right) }$   are analytical and compatible solutions of the relaxation operator denoted at  Eq. (A4a) of Ref. \cite{tsoref1996}. The relaxation rates  $R_{1}^{\left( 2\right) }$, $R_{2}^{\left( 2\right) }$ and $R_{5}^{\left( 2\right) }$  are the new ones introduced to describe the dynamics of $2$th coherence order density matrix elements.

 One of the main characteristics on the solution of the 2nd coherence order elements points out on the transformation matrix which explicitly depend on the multiplication of two transformations, one depending only on constant elements and the other one depending on the spectral density functions.

The other one characteristic arises on the transformation that depends on the cartesian parameters, highlighting its $6\times6$ dimension, which apparently is hard to diagonalize analytically. In this particular case, and by algebraic procedures, it was possible to decompose on two transformations with $3\times3$ dimensions each,  which is clearly observed at its matrix notation. From the fundamentals of algebra, there are some procedures to solve linear systems of low orders, as $3\times3$ dimensions, and they were used to solve them. By similar arguments were found the solutions of the 3rd coherence order.

\textit{First coherences order:} The density matrix elements of $1$th order  denoted as $\rho _{2,1}\left( t\right)$,  $\rho _{3,2}\left( t\right)$,  $\rho _{4,3}\left( t\right)$, $\rho _{5,4}\left( t\right)$, $\rho _{6,5}\left( t\right)$,  $\rho _{7,6}\left( t\right)$, and  $\rho _{8,7}\left( t\right)$  can be represented with the Eq. (\ref{RedfieldElementoGeral}) and (\ref{RedfieldElementoGeralInitialConditions}) such that the subscript indices  $p$, $n$ and $q$ obey the values $p \equiv \left\{ 1 , 2 , 3 , 4 , 5 , 6 , 7\right\}$,  $n \equiv \left\{ 1 , 2 , 3 , 4 , 5 , 6 , 7 \right\}$ and the coherence order $q=1$. With this notation in hands, the transformation  $\mathbf{W}^{\left( 1\right)}$ is the multiplication of the transformation $\mathbf{V}^{\left( 1\right)}$ with elements $V_{j,k}^{\left( 1\right)}$ and the transformation  $\mathbf{U}^{\left( 1\right)} $ with constant elements $U_{j,k}^{\left( 1\right)}$. Both of them are  denoted  $\mathbf{W}^{\left( 1\right)}=\mathbf{V}^{\left( 1\right)}\mathbf{U}^{\left( 1\right)}$ and in matrix notation by
\begin{equation*}
\mathbf{W}^{\left( 1\right)}=\left[ 
\begin{array}{ccccccc}
V_{1,1}^{\left( 1\right)} & V_{1,2}^{\left( 1\right)} & V_{1,3}^{\left( 1\right)} & V_{1,4}^{\left( 1\right)} & V_{1,5}^{\left( 1\right)} & V_{1,6}^{\left( 1\right)} & V_{1,7}^{\left( 1\right)} \\ 
V_{2,1}^{\left( 1\right)} & V_{2,2}^{\left( 1\right)} & V_{2,3}^{\left( 1\right)} & V_{2,4}^{\left( 1\right)} & V_{2,5}^{\left( 1\right)} & V_{2,6}^{\left( 1\right)} & V_{2,7}^{\left( 1\right)} \\ 
V_{3,1}^{\left( 1\right)} & V_{3,2}^{\left( 1\right)} & V_{3,3}^{\left( 1\right)} & V_{3,4}^{\left( 1\right)} & V_{3,5}^{\left( 1\right)} & V_{3,6}^{\left( 1\right)} & V_{3,7}^{\left( 1\right)} \\ 
V_{4,1}^{\left( 1\right)} & V_{4,2}^{\left( 1\right)} & V_{4,3}^{\left( 1\right)} & V_{4,4}^{\left( 1\right)} & V_{4,5}^{\left( 1\right)} & V_{4,6}^{\left( 1\right)} & V_{4,7}^{\left( 1\right)} \\ 
V_{5,1}^{\left( 1\right)} & V_{5,2}^{\left( 1\right)} & V_{5,3}^{\left( 1\right)} & V_{5,4}^{\left( 1\right)} & V_{5,5}^{\left( 1\right)} & V_{5,6}^{\left( 1\right)} & V_{5,7}^{\left( 1\right)} \\ 
V_{6,1}^{\left( 1\right)} & V_{6,2}^{\left( 1\right)} & V_{6,3}^{\left( 1\right)} & V_{6,4}^{\left( 1\right)} & V_{6,5}^{\left( 1\right)} & V_{6,6}^{\left( 1\right)} & V_{6,7}^{\left( 1\right)} \\ 
V_{7,1}^{\left( 1\right)} & V_{7,2}^{\left( 1\right)} & V_{7,3}^{\left( 1\right)} & V_{7,4}^{\left( 1\right)} & V_{7,5}^{\left( 1\right)} & V_{7,6}^{\left( 1\right)} & V_{7,7}^{\left( 1\right)}
\end{array}%
\right] \left[ 
\begin{array}{ccccccc}
\frac{\sqrt{7}}{\sqrt{22}} & \frac{1}{\sqrt{66}} & -\frac{\sqrt{5}}{\sqrt{66}%
} & -\frac{\sqrt{2}}{\sqrt{11}} & -\frac{\sqrt{5}}{\sqrt{66}} & \frac{1}{%
\sqrt{66}} & \frac{\sqrt{7}}{\sqrt{22}} \\ 
\frac{-7}{\sqrt{21}} & -1 & 0 & \frac{2}{\sqrt{3}} & \sqrt{5} & 3 & \frac{14%
}{\sqrt{21}} \\ 
-1 & -\frac{2}{\sqrt{21}} & \frac{\sqrt{5}}{\sqrt{21}} & \frac{4}{\sqrt{7}}
& \frac{5\sqrt{5}}{\sqrt{21}} & \frac{14}{\sqrt{21}} & 3 \\ 
-\frac{1}{\sqrt{44}} & \frac{\sqrt{7}}{\sqrt{33}} & -\frac{\sqrt{35}}{2\sqrt{%
33}} & 0 & \frac{\sqrt{35}}{2\sqrt{33}} & -\frac{\sqrt{7}}{\sqrt{33}} & 
\frac{1}{\sqrt{44}} \\ 
-\frac{\sqrt{5}}{\sqrt{22}} & \frac{\sqrt{15}}{\sqrt{154}} & \frac{3\sqrt{3}%
}{\sqrt{154}} & 0 & -\frac{3\sqrt{3}}{\sqrt{154}} & -\frac{\sqrt{15}}{\sqrt{%
154}} & \frac{\sqrt{5}}{\sqrt{22}} \\ 
\frac{1}{\sqrt{429}} & -\frac{\sqrt{21}}{\sqrt{429}} & \frac{\sqrt{35}}{%
\sqrt{143}} & -\frac{35}{\sqrt{7}\sqrt{429}} & \frac{\sqrt{35}}{\sqrt{143}}
& -\frac{\sqrt{7}}{\sqrt{143}} & \frac{1}{\sqrt{429}} \\ 
\frac{\sqrt{5}}{2\sqrt{13}} & -\frac{4\sqrt{5}}{\sqrt{273}} & \frac{1}{2%
\sqrt{273}} & \frac{2\sqrt{15}}{\sqrt{273}} & \frac{1}{2\sqrt{273}} & -\frac{%
4\sqrt{5}}{\sqrt{273}} & \frac{\sqrt{5}}{2\sqrt{13}}%
\end{array}%
\right] \text{,}
\end{equation*}
and the inverse transformation $\overline{\mathbf{W}}^{\left( 1\right)} =\overline{\mathbf{U}}^{\left( 1\right)} \overline{\mathbf{V}}^{\left( 1\right)}$  denoted by
\begin{equation*}
\overline{\mathbf{W}}^{\left( 1\right)}=\left[ 
\begin{array}{ccccccc}
\frac{\sqrt{21}}{\sqrt{66}} & -\frac{5\sqrt{7}}{\sqrt{588}} & \frac{\sqrt{21}%
}{\sqrt{84}} & -\frac{\sqrt{3}}{\sqrt{132}} & -\frac{\sqrt{175}}{\sqrt{770}}
& \frac{1}{\sqrt{429}} & \frac{\sqrt{5}}{\sqrt{52}} \\ 
\frac{1}{\sqrt{66}} & -\frac{8\sqrt{3}}{\sqrt{588}} & \frac{5}{\sqrt{84}} & 
\frac{2\sqrt{7}}{\sqrt{132}} & \frac{\sqrt{75}}{\sqrt{770}} & \frac{-\sqrt{21%
}}{\sqrt{429}} & -\frac{8\sqrt{5}}{\sqrt{1092}} \\ 
-\frac{\sqrt{5}}{\sqrt{66}} & -\frac{3\sqrt{15}}{\sqrt{588}} & \frac{2\sqrt{5%
}}{\sqrt{84}} & -\frac{\sqrt{35}}{\sqrt{132}} & \frac{3\sqrt{15}}{\sqrt{770}}
& \frac{\sqrt{105}}{\sqrt{429}} & \frac{1}{\sqrt{1092}} \\ 
-\frac{\sqrt{12}}{\sqrt{66}} & -\frac{8}{\sqrt{588}} & \frac{2\sqrt{3}}{%
\sqrt{84}} & 0 & 0 & \frac{-5\sqrt{7}}{\sqrt{429}} & \frac{4\sqrt{15}}{\sqrt{%
1092}} \\ 
-\frac{\sqrt{5}}{\sqrt{66}} & -\frac{\sqrt{15}}{\sqrt{588}} & \frac{\sqrt{5}%
}{\sqrt{84}} & \frac{\sqrt{35}}{\sqrt{132}} & -\frac{3\sqrt{15}}{\sqrt{770}}
& \frac{\sqrt{105}}{\sqrt{429}} & \frac{1}{\sqrt{1092}} \\ 
\frac{1}{\sqrt{66}} & 0 & \frac{1}{\sqrt{84}} & -\frac{2\sqrt{7}}{\sqrt{132}}
& -\frac{\sqrt{75}}{\sqrt{770}} & \frac{-\sqrt{21}}{\sqrt{429}} & -\frac{8%
\sqrt{5}}{\sqrt{1092}} \\ 
\frac{\sqrt{21}}{\sqrt{66}} & \frac{\sqrt{7}}{\sqrt{588}} & 0 & \frac{\sqrt{3%
}}{\sqrt{132}} & \frac{\sqrt{175}}{\sqrt{770}} & \frac{1}{\sqrt{429}} & 
\frac{\sqrt{5}}{\sqrt{52}}%
\end{array}%
\right] \left[ 
\begin{array}{ccccccc}
 \overline{V}_{1,1}^{\left( 1\right)} & \overline{V}_{1,2}^{\left( 1\right)} &  \overline{V}_{1,3}^{\left( 1\right)} &  \overline{V}_{1,4}^{\left( 1\right)}
&  \overline{V}_{1,5}^{\left( 1\right)} &  \overline{V}_{1,6}^{\left( 1\right)}&  \overline{V}_{1,7}^{\left( 1\right)} \\ 
 \overline{V}_{2,1}^{\left( 1\right)} &  \overline{V}_{2,2}^{\left( 1\right)} &  \overline{V}_{2,3}^{\left( 1\right)} &  \overline{V}_{2,4}^{\left( 1\right)}
&  \overline{V}_{2,5}^{\left( 1\right)} &  \overline{V}_{2,6}^{\left( 1\right)}&  \overline{V}_{2,7}^{\left( 1\right)} \\ 
 \overline{V}_{3,1}^{\left( 1\right)} &  \overline{V}_{3,2}^{\left( 1\right)} &  \overline{V}_{3,3}^{\left( 1\right)} &  \overline{V}_{3,4}^{\left( 1\right)}
&  \overline{V}_{3,5}^{\left( 1\right)}&  \overline{V}_{3,6}^{\left( 1\right)} &  \overline{V}_{3,7}^{\left( 1\right)} \\ 
 \overline{V}_{4,1}^{\left( 1\right)} &  \overline{V}_{4,2}^{\left( 1\right)} &  \overline{V}_{4,3}^{\left( 1\right)} &  \overline{V}_{4,4}^{\left( 1\right)}
&  \overline{V}_{4,5}^{\left( 1\right)} &  \overline{V}_{4,6}^{\left( 1\right)}&  \overline{V}_{4,7}^{\left( 1\right)} \\ 
 \overline{V}_{5,1}^{\left( 1\right)} &  \overline{V}_{5,2}^{\left( 1\right)}&  \overline{V}_{5,3}^{\left( 1\right)} &  \overline{V}_{5,4}^{\left( 1\right)}
&  \overline{V}_{5,5}^{\left( 1\right)} &  \overline{V}_{5,6}^{\left( 1\right)} &  \overline{V}_{5,7}^{\left( 1\right)} \\ 
 \overline{V}_{6,1}^{\left( 1\right)} &  \overline{V}_{6,2}^{\left( 1\right)}&  \overline{V}_{6,3}^{\left( 1\right)} &  \overline{V}_{6,4}^{\left( 1\right)}
&  \overline{V}_{6,5}^{\left( 1\right)}&  \overline{V}_{6,6}^{\left( 1\right)} &  \overline{V}_{6,7}^{\left( 1\right)} \\ 
 \overline{V}_{7,1}^{\left( 1\right)} &  \overline{V}_{7,2}^{\left( 1\right)} &  \overline{V}_{7,3}^{\left( 1\right)} &  \overline{V}_{7,4}^{\left( 1\right)}
&  \overline{V}_{7,5}^{\left( 1\right)} &  \overline{V}_{7,6}^{\left( 1\right)} &  \overline{V}_{7,7}^{\left( 1\right)}
\end{array}%
\right] \text{.}
\end{equation*}
The transformation matrix $\mathbf{V}^{\left( 1\right)}$ is used to diagonalize the  Redfield's superoperator $\boldsymbol{\mathcal{J}}^{\left( 1\right)}$ such that each $ {\mathcal{J}}^{\left( 1\right)}_{j,k}$  matrix element is resumed in Tab. \ref{tab:Js1raOrdem}. The diagonalization procedure  obeys the mathematical equation $\boldsymbol{\mathcal{D}}^{\left( 1\right)} = \mathbf{V}^{\left( 1\right)} \boldsymbol{\mathcal{J}}^{\left( 1\right)}\overline{\mathbf{ V}}^{\left( 1\right)}$ and the eigenvalues are denoted by $\lambda_{k}^{\left( 1\right)} $ which are elements of the diagonal matrix $\boldsymbol{\mathcal{D}}^{\left( 1\right)}$.

At this stage, the main difficulty to solve analytically the linear system is the nonviable decomposition on transformations with lower dimensions, because the dimensionality can not always be broken into other linear systems of smaller dimensionality. This characteristic was highlighted on spins systems with $I>2$ and inevitably the analysis must be accomplished by numerical procedures \cite{kelly1992}.    For this reason, a general notation is considered but the linear system will be used with an experimental implementation as it is explored in Sec. \ref{sec:ExperimentalDescriptionOn133Cs}.

\begin{table}[h!]
\begin{center}
\begin{tabular}{|c|c|c|c|c|} \hline \hline
   \multicolumn{5}{|c|}{$\mathcal{J}_{j,k}^{\left( 1\right)}$}	\\ \hline
&   \multicolumn{4}{|c|}{$k$}	\\ \cline{2-5}
$j$ & 1 &  2 & 3 &   	\\  \hline
1 & $-6{J}_{0}-\frac{55}{3}{J}_{1}-\frac{77}{3}{J}_{2}$ & $\frac{2}{7}\left( {J}_{0}-{J}_{2}\right) \sqrt{66}$ & $-\frac{3}{14}\left( {J}_{0}-{J}_{2}\right) \sqrt{154}$ & $0$  	\\
2 & $-\left( {J}_{0}-{J}_{2}\right) \sqrt{66}$ & $-\frac{87}{7}{J}_{0}-\frac{83}{7}{J}_{1}-\frac{278}{7}%
{J}_{2}$ & $\left( \frac{11}{7}{J}_{0}+\frac{8}{7}{J}_{1}+\frac{44}{%
7}{J}_{2}\right) \sqrt{21}$ & $0$  	\\
3 & $-\frac{6}{7}\left( {J}_{0}-{J}_{2}\right) \sqrt{154}$ & $-\left( \frac{88}{7}{J}_{0}+\frac{64}{7}{J}_{1}+\frac{352}{7}{J}_{2}\right) \frac{\sqrt{3}}{\sqrt{7}}$ & $\frac{23}{7}{J}_{0}-\frac{3}{7}{J}_{1}+\frac{162}{7}%
{J}_{2}$ & $0$  	\\
4 & $0$ & $0$ & $0$ & $0$  	\\
5 & $0$ & $-\left( \frac{2}{7}{J}_{0}+\frac{4}{7}{J}_{1}-\frac{6}{7}{J}_{2}\right) \frac{\sqrt{330}}{\sqrt{7}}$ & $\left( \frac{1}{7}{J}_{0}+\frac{2}{7}{J}_{1}-\frac{3}{7}%
{J}_{2}\right) \sqrt{110}$ & $0$  	\\
6 & $\left( \frac{20}{429}{J}_{1}\right) \sqrt{546}$ & $\left( \frac{80}{3003}{J}_{1}\right) \sqrt{1001}$  & $\left( -\frac{20}{429}{J}_{1}\right) \sqrt{429}$ &  $0$  	\\
7 & $-\left( \frac{2}{7}{J}_{0}+\frac{4}{3}{J}_{1}-\frac{34}{21}{J}_{2}\right) \frac{\sqrt{910}}{\sqrt{11}}$ & $ 0$  & $0$ &  $0$	\\ \hline \hline 
& 4 &  5 & 6 & 7 	\\  \hline
1 & $0$ & $0$ & $0$ &  $-\left( \frac{2}{7}{J}_{0}+\frac{4}{3}{J}_{1}-\frac{34}{21}{J}_{2}\right) \frac{\sqrt{910}}{\sqrt{11}}$  	\\
2 & $0$ & $-\frac{2}{7}\left( {J}_{0}+2{J}_{1}-3{J}_{2}\right) \sqrt{2310}$ & $0$ &  $0$  	\\
3 & $0$ & $-\frac{8}{7}\left( {J}_{0}+2{J}_{1}-3{J}_{2}\right) \sqrt{110}$ & $0$ &  $0$ 		\\
4 & $-\frac{29}{11}{J}_{0}-\frac{375}{11}{J}_{1}-\frac{366}{%
11}{J}_{2}$ & $-\frac{7}{11}\left( {J}_{0}+8{J}_{1}-9{J}_{2}\right) 
\sqrt{10}$ & $0$ &  $0$  	\\
5 & $-\frac{7}{11}\left( {J}_{0}+8{J}_{1}-9{J}_{2}\right) 
\sqrt{10}$ & $-\frac{402}{77}{J}_{0}-\frac{2281}{77}{J}_{1}-\frac{2707%
}{77}{J}_{2}$ & $0$ &  $0$  	\\
6 & $0$ & $0$  & $-\frac{12}{13}{J}_{0}-\frac{6197}{429}{J}_{1}-11{J}_{2}$ &  $-\left( \frac{7}{13}{J}_{0}+\frac{272}{39} {J}_{1}-7{J}_{2}\right) \frac{\sqrt{60}}{\sqrt{11}}$  	\\
7 & $0$ & $0$  & $-\left( \frac{14}{13}{J}_{0}+\frac{168}{13}{J}_{1}-%
{14}{J}_{2}\right) \frac{\sqrt{15}}{\sqrt{11}}$ &  $-\frac{53}{13}{J}_{0}-\frac{1427}{39}{J}_{1}-\frac{118}{3}{J}_{2}$	\\ \hline \hline
\end{tabular}
\end{center}
\caption{Matrix elements of Redfield's superoperator  $\boldsymbol{\mathcal{J}}^{\left( 1\right)}$ used to define the dynamics of the first coherence orders of the density matrix elements $  {{\rho }_{k+1,k}}  \left( t\right)$ in terms of the spectral density functions ${J}_{0}$, ${J}_{1}$  and ${J}_{2}$.} \label{tab:Js1raOrdem}
\end{table}

Therefore, applying the general notation of Eq. (\ref{RedfieldElementoGeral}), the first order density matrix elements are denoted by
\begin{equation}
\rho _{1+n,n}\left( t\right)   = \sum_{p=1}^{7}\overline{W}_{n,p}^{\left( 1\right)}\exp \left[ -R_{p}^{\left( 1 \right) }\left( t-t_{0}\right) \right] \widetilde{\rho }_{p}\left( t_{0}\right)  \text{,}
\end{equation}
where $R_{p}^{\left( 1 \right) } = - C \lambda_{p}^{\left( 1\right)} $ represent the relaxation rates that characterize the dynamics of the first order density matrix elements.

\textit{Zero coherences order:} the density matrix elements of $0$th order   denoted as  $\rho _{1,1}\left( t\right)$,  $\rho _{2,2}\left( t\right)$,  $\rho _{3,3}\left( t\right)$, $\rho _{4,4}\left( t\right)$, $\rho _{5,5}\left( t\right)$,  $\rho _{6,6}\left( t\right)$,  $\rho _{7,7}\left( t\right)$, and  $\rho _{8,8}\left( t\right)$  can be represented with the Eq. (\ref{RedfieldElementoGeral}) and (\ref{RedfieldElementoGeralInitialConditions}) such that the subscript indices  $p$, $n$ and $q$ obey the values $p \equiv \left\{ 1 , 2 , 3 , 4 , 5 , 6 , 7 , 8 \right\}$,  $n \equiv \left\{ 1 , 2 , 3 , 4 , 5 , 6 , 7 , 8 \right\}$ and the coherence order $q=0$. With this notation in hands, the transformation  $\mathbf{W}^{\left( 0\right)}$ is the multiplication of the transformation  $\mathbf{V}^{\left( 0\right)}$  with elements $V_{j,k}^{\left( 0\right)}$ and the transformation  $\mathbf{U}^{\left( 0\right)}$ with constant elements $U_{j,k}^{\left( 0\right)}$, such that the transformation $\mathbf{W}^{\left( 0\right)}=\mathbf{V}^{\left( 0\right)}\mathbf{U}^{\left( 0\right)}$ is denoted by
\begin{small}
\begin{equation*}
\mathbf{W}^{\left( 0\right)} =\left[ 
\begin{array}{cccccccc}
V_{1,1}^{\left( 0\right)} & V_{1,2}^{\left( 0\right)} & V_{1,3}^{\left( 0\right)} & V_{1,4}^{\left( 0\right)} & V_{1,5}^{\left( 0\right)} & V_{1,6} ^{\left( 0\right)}& V_{1,7}^{\left( 0\right)} & V_{1,8}^{\left( 0\right)}
\\ 
V_{2,1}^{\left( 0\right)} & V_{2,2}^{\left( 0\right)} & V_{2,3}^{\left( 0\right)} & V_{2,4}^{\left( 0\right)} & V_{2,5}^{\left( 0\right)} & V_{2,6}^{\left( 0\right)} & V_{2,7}^{\left( 0\right)} & V_{2,8}^{\left( 0\right)}
\\ 
V_{3,1}^{\left( 0\right)} & V_{3,2}^{\left( 0\right)} & V_{3,3}^{\left( 0\right)} & V_{3,4}^{\left( 0\right)} & V_{3,5}^{\left( 0\right)} & V_{3,6}^{\left( 0\right)} & V_{3,7}^{\left( 0\right)} & V_{3,8}^{\left( 0\right)}
\\ 
V_{4,1}^{\left( 0\right)} & V_{4,2}^{\left( 0\right)} & V_{4,3}^{\left( 0\right)} & V_{4,4}^{\left( 0\right)} & V_{4,5}^{\left( 0\right)} & V_{4,6}^{\left( 0\right)} & V_{4,7}^{\left( 0\right)} & V_{4,8}^{\left( 0\right)}
\\ 
V_{5,1}^{\left( 0\right)} & V_{5,2}^{\left( 0\right)} & V_{5,3}^{\left( 0\right)} & V_{5,4}^{\left( 0\right)} & V_{5,5}^{\left( 0\right)} & V_{5,6}^{\left( 0\right)} & V_{5,7}^{\left( 0\right)} & V_{5,8}^{\left( 0\right)}
\\ 
V_{6,1}^{\left( 0\right)} & V_{6,2}^{\left( 0\right)} & V_{6,3}^{\left( 0\right)} & V_{6,4}^{\left( 0\right)} & V_{6,5}^{\left( 0\right)} & V_{6,6}^{\left( 0\right)} & V_{6,7}^{\left( 0\right)} & V_{6,8}^{\left( 0\right)}
\\ 
V_{7,1}^{\left( 0\right)} & V_{7,2}^{\left( 0\right)} & V_{7,3}^{\left( 0\right)} & V_{7,4}^{\left( 0\right)} & V_{7,5}^{\left( 0\right)} & V_{7,6}^{\left( 0\right)} & V_{7,7}^{\left( 0\right)} & V_{7,8}^{\left( 0\right)}
\\ 
V_{8,1}^{\left( 0\right)} & V_{8,2}^{\left( 0\right)} & V_{8,3}^{\left( 0\right)} & V_{8,4}^{\left( 0\right)} & V_{8,5}^{\left( 0\right)} & V_{8,6}^{\left( 0\right)} & V_{8,7}^{\left( 0\right)} & V_{8,8}^{\left( 0\right)}
\end{array}%
\right] \left[ 
\begin{array}{cccccccc}
\frac{1}{\sqrt{8}} & \frac{1}{\sqrt{8}} & \frac{1}{\sqrt{8}} & \frac{1}{%
\sqrt{8}} & \frac{1}{\sqrt{8}} & \frac{1}{\sqrt{8}} & \frac{1}{\sqrt{8}} & 
\frac{1}{\sqrt{8}} \\ 
-\frac{14}{\sqrt{336}} & -\frac{12}{\sqrt{336}} & \frac{-9}{\sqrt{336}} & 
\frac{-5}{\sqrt{336}} & 0 & \frac{6}{\sqrt{336}} & \frac{13}{\sqrt{336}} & 
\frac{21}{\sqrt{336}} \\ 
0 & -\frac{26}{\sqrt{4368}} & -\frac{39}{\sqrt{4368}} & -\frac{39}{\sqrt{4368%
}} & -\frac{26}{\sqrt{4368}} & 0 & \frac{39}{\sqrt{4368}} & \frac{91}{\sqrt{%
4368}} \\ 
-\frac{24}{\sqrt{1188}} & \frac{15}{\sqrt{1188}} & \frac{29}{\sqrt{1188}} & 
\frac{8}{\sqrt{1188}} & -\frac{13}{\sqrt{1188}} & -\frac{20}{\sqrt{1188}} & -%
\frac{20}{\sqrt{1188}} & \frac{25}{\sqrt{1188}} \\ 
\frac{15}{\sqrt{1188}} & 0 & -\frac{40}{\sqrt{1188}} & \frac{5}{\sqrt{1188}}
& \frac{20}{\sqrt{1188}} & -\frac{5}{\sqrt{1188}} & \frac{25}{\sqrt{1188}} & 
-\frac{20}{\sqrt{1188}} \\ 
\frac{7}{\sqrt{616}} & -\frac{13}{\sqrt{616}} & -\frac{3}{\sqrt{616}} & 
\frac{9}{\sqrt{616}} & \frac{9}{\sqrt{616}} & -\frac{3}{\sqrt{616}} & -\frac{%
13}{\sqrt{616}} & \frac{7}{\sqrt{616}} \\ 
-\frac{7}{\sqrt{2184}} & \frac{23}{\sqrt{2184}} & -\frac{17}{\sqrt{2184}} & -%
\frac{15}{\sqrt{2184}} & \frac{15}{\sqrt{2184}} & \frac{17}{\sqrt{2184}} & -%
\frac{23}{\sqrt{2184}} & \frac{7}{\sqrt{2184}} \\ 
-\frac{1}{\sqrt{3432}} & \frac{7}{\sqrt{3432}} & -\frac{21}{\sqrt{3432}} & 
\frac{35}{\sqrt{3432}} & -\frac{35}{\sqrt{3432}} & \frac{21}{\sqrt{3432}} & -%
\frac{7}{\sqrt{3432}} & \frac{1}{\sqrt{3432}}%
\end{array}%
\right] \text{,}
\end{equation*}
\end{small}
and the inverse transformation $\overline{\mathbf{W}}^{\left( 0\right)}  =\overline{\mathbf{U}}^{\left( 0\right)} \overline{\mathbf{V}}^{\left( 0\right)} $  denoted by 

\begin{small}
\begin{equation*}
\overline{\mathbf{W}}^{\left( 0\right)}     =   \left[ 
\begin{array}{cccccccc}
\frac{1}{\sqrt{8}} & -\frac{7}{\sqrt{84}} & \frac{21}{\sqrt{1092}} & -\frac{4%
}{\sqrt{132}} & -\frac{7}{\sqrt{3300}} & \frac{7}{\sqrt{616}} & -\frac{7}{%
\sqrt{2184}} & -\frac{1}{\sqrt{3432}} \\ 
\frac{1}{\sqrt{8}} & -\frac{3}{\sqrt{84}} & \frac{5}{\sqrt{1092}} & \frac{5}{%
\sqrt{132}} & \frac{20}{\sqrt{3300}} & -\frac{13}{\sqrt{616}} & \frac{23}{%
\sqrt{2184}} & \frac{7}{\sqrt{3432}} \\ 
\frac{1}{\sqrt{8}} & 0 & -\frac{6}{\sqrt{1092}} & -\frac{1}{\sqrt{132}} & -%
\frac{28}{\sqrt{3300}} & -\frac{3}{\sqrt{616}} & -\frac{17}{\sqrt{2184}} & -%
\frac{21}{\sqrt{3432}} \\ 
\frac{1}{\sqrt{8}} & \frac{2}{\sqrt{84}} & -\frac{12}{\sqrt{1092}} & \frac{4%
}{\sqrt{132}} & \frac{19}{\sqrt{3300}} & \frac{9}{\sqrt{616}} & -\frac{15}{%
\sqrt{2184}} & \frac{35}{\sqrt{3432}} \\ 
\frac{1}{\sqrt{8}} & \frac{3}{\sqrt{84}} & -\frac{13}{\sqrt{1092}} & \frac{1%
}{\sqrt{132}} & \frac{16}{\sqrt{3300}} & \frac{9}{\sqrt{616}} & \frac{15}{%
\sqrt{2184}} & -\frac{35}{\sqrt{3432}} \\ 
\frac{1}{\sqrt{8}} & \frac{3}{\sqrt{84}} & -\frac{9}{\sqrt{1092}} & -\frac{8%
}{\sqrt{132}} & -\frac{35}{\sqrt{3300}} & -\frac{3}{\sqrt{616}} & \frac{17}{%
\sqrt{2184}} & \frac{21}{\sqrt{3432}} \\ 
\frac{1}{\sqrt{8}} & \frac{2}{\sqrt{84}} & 0 & 0 & \frac{15}{\sqrt{3300}} & -%
\frac{13}{\sqrt{616}} & -\frac{23}{\sqrt{2184}} & -\frac{7}{\sqrt{3432}} \\ 
\frac{1}{\sqrt{8}} & 0 & \frac{14}{\sqrt{1092}} & \frac{3}{\sqrt{132}} & 0 & 
\frac{7}{\sqrt{616}} & \frac{7}{\sqrt{2184}} & \frac{1}{\sqrt{3432}}%
\end{array}%
\right] \left[ 
\begin{array}{cccccccc}
\overline{V}_{1,1}  & \overline{V}_{1,2}  & \overline{V}_{1,3}  & \overline{V}_{1,4} 
& \overline{V}_{1,5}  & \overline{V}_{1,6}  & \overline{V}_{1,7}  & 
\overline{V}_{1,8}  \\ 
\overline{V}_{2,1}  & \overline{V}_{2,2}  & \overline{V}_{2,3}  & \overline{V}_{2,4} 
& \overline{V}_{2,5}  & \overline{V}_{2,6}  & \overline{V}_{2,7}  & 
\overline{V}_{2,8}  \\ 
\overline{V}_{3,1}  & \overline{V}_{3,2}  & \overline{V}_{3,3}  & \overline{V}_{3,4} 
& \overline{V}_{3,5}  & \overline{V}_{3,6}  & \overline{V}_{3,7}  & 
\overline{V}_{3,8}  \\ 
\overline{V}_{4,1}  & \overline{V}_{4,2}  & \overline{V}_{4,3}  & \overline{V}_{4,4} 
& \overline{V}_{4,5}  & \overline{V}_{4,6}  & \overline{V}_{4,7}  & 
\overline{V}_{4,8}  \\ 
\overline{V}_{5,1} & \overline{V}_{5,2}  & \overline{V}_{5,3}  & \overline{V}_{5,4} 
& \overline{V}_{5,5}  & \overline{V}_{5,6}  & \overline{V}_{5,7}  & 
\overline{V}_{5,8}  \\ 
\overline{V}_{6,1}  & \overline{V}_{6,2}  & \overline{V}_{6,3}  & \overline{V}_{6,4} 
& \overline{V}_{6,5}  & \overline{V}_{6,6}  & \overline{V}_{6,7}  & 
\overline{V}_{6,8}  \\ 
\overline{V}_{7,1}  & \overline{V}_{7,2}  & \overline{V}_{7,3}  & \overline{V}_{7,4} 
& \overline{V}_{7,5}  & \overline{V}_{7,6}  & \overline{V}_{7,7}  & 
\overline{V}_{7,8}  \\ 
\overline{V}_{8,1}  & \overline{V}_{8,2}  & \overline{V}_{8,3}  & \overline{V}_{8,4} 
& \overline{V}_{8,5}  & \overline{V}_{8,6}  & \overline{V}_{8,7}  & 
\overline{V}_{8,8} %
\end{array}%
\right] \text{.}
\end{equation*}
\end{small}

\begin{table}[t]
\begin{center}
\begin{tabular}{|c|c|c|c|c|} \hline \hline
   \multicolumn{5}{|c|}{$\mathcal{J}_{j,k}^{\left( 0\right)}$}	\\ \hline
&   \multicolumn{4}{|c|}{$k$}	\\ \cline{2-5}
$j$ & 1 &  2 & 3 &  4 	\\  \hline
1 & $0$ & $0$  & $0$ &  $0$  	\\
2 & $0$ & $\frac{8}{14}{J}_{1}-\frac{85}{14}{J}_{2}$  & $-\frac{1}{182}\left( 180{J}_{1}+135{J}_{2}\right) \sqrt{13}$ &  $-\frac{5}{7}\left( {J}_{1}-{J}_{2}\right) \sqrt{77}$  	\\
3 & $0$ & $\frac{1}{14}\left( 36{J}_{1}+27{J}_{2}\right) \sqrt{13}$  & $-\frac{208}{14}{J}_{1}-\frac{247}{14}{J}_{2}$ &  $-\frac{1}{7}\left( {J}_{1}-{J}_{2}\right) \sqrt{1001}$  	\\
4 & $0$ & $-\frac{2}{3}\left( {J}_{1}-{J}_{2}\right) \sqrt{77}$  & $\frac{2}{39}\left({J}_{1}-{J}_{2}\right) \sqrt{1001}$ &  $-\frac{721}{33}{J}_{1}-\frac{273}{11}{J}_{2}$  	\\
5 & $0$ & $\frac{10}{21}\left({J}_{1}-{J}_{2}\right) \sqrt{77}$  & $-\frac{10}{273}\left({J}_{1}-{J}_{2}\right) \sqrt{1001}$ &  $-\frac{355}{33}{J}_{1}-\frac{65}{11}{J}_{2}$  	\\
6 & $0$ & $\frac{20}{21}\left({J}_{1}-{J}_{2}\right) \sqrt{66}$  & $-\frac{100}{273}\left( {J}_{1}-{J}_{2}\right) \sqrt{858}$ &  $\frac{70}{33}\left({J}_{1}-{J}_{2}\right) \sqrt{42}$  	\\
7 & $0$ & $0$  & $0$ &  $-\frac{20}{77}\left( {J}_{1}-{J}_{2}\right) \sqrt{2002}$	\\
8 & $0$ & $0$  & $0$ &  $0$	\\ \hline \hline 
& 5 &  6 & 7 & 8 	\\  \hline
1 & $0$ & $0$  & $0$ &  $0$  	\\
2 & $-\frac{1}{7}\left({J}_{1}-{J}_{2}\right) \sqrt{77}$ & $-\frac{10}{21}\left({J}_{1}-{J}_{2}\right) \sqrt{66}$  & $0$ &  $0$  	\\
3 & $-\frac{1}{35}\left({J}_{1}-{J}_{2}\right) \sqrt{1001}$ & $-\frac{10}{21}\left({J}_{1}-{J}_{2}\right) \sqrt{858}$  & $0$ &  $0$ 		\\
4 & $\frac{497}{165}{J}_{1}+\frac{91}{55}{J}_{2}$ & $-\frac{70}{99}\left({J}_{1}-{J}_{2}\right) \sqrt{42}$  & $-\frac{20}{33}\left({J}_{1}-{J}_{2}\right) \sqrt{2002}$ &  $0$  	\\
5 & $-\frac{1289}{33}{J}_{1}-\frac{377}{11}{J}_{2}$ & $\frac{350}{99}\left({J}_{1}-{J}_{2}\right) \sqrt{42}$  & $\frac{100}{231}\left({J}_{1}-{J}_{2}\right) \sqrt{2002}$ &  $0$  	\\
6 & $\frac{98}{33}\left({J}_{1}-{J}_{2}\right) \sqrt{42}$ & $-\frac{2680}{77}{J}_{1}-\frac{2710}{77}{J}_{2}$  & $0$ &  $0$  	\\
7 & $-\frac{4}{77}\left({J}_{1}-{J}_{2}\right) \sqrt{2002}$ & $0$  & $-\frac{530}{13}{J}_{1}-\frac{510}{13}{J}_{2}$ &  $-\frac{280}{143}\left({J}_{1}-{J}_{2}\right) \sqrt{77} $	\\
8 & $0$ & $0$  & $-\frac{280}{143}\left( {J}_{1}-{J}_{2}\right) \sqrt{77}$ &  $-\frac{224}{13}{J}_{1}-\frac{140}{13}{J}_{2}$	\\ \hline \hline
\end{tabular}
\end{center}
\caption{Matrix elements of the Redfield's superoperator  $\boldsymbol{\mathcal{J}}^{\left( 0\right)}$ used to define the dynamics of the zero coherence orders of the density matrix elements $  {{\rho }_{k,k}}  \left( t\right)$ in terms of the spectral density functions ${J}_{1}$ and ${J}_{2}$. }
\label{tab:Js0maOrdem}
\end{table}

The transformation  $\mathbf{V}^{\left( 0\right)}$ is used to diagonalize Redfield's superoperator $\boldsymbol{\mathcal{J}}^{\left( 0\right)}$ such that each $ {\mathcal{J}}^{\left( 0\right)}_{j,k}$  matrix element is resumed in Tab. \ref{tab:Js0maOrdem}. The diagonalization procedure   obeys the mathematical equation $\boldsymbol{\mathcal{D}}^{\left( 0\right)} = \mathbf{V}^{\left( 0\right)} \boldsymbol{\mathcal{J}}^{\left( 0\right)}\overline{\mathbf{V}}^{\left( 0\right)}$ and the eigenvalues are denoted by $\lambda_{k}^{\left( 0\right)} $, which are elements of the diagonal transformation $\boldsymbol{\mathcal{D}}^{\left( 0\right)}$.

Therefore, applying the general notation of Eq. (\ref{RedfieldElementoGeral}), the zero order density matrix elements are denoted by
\begin{equation}
\rho _{0+n,n}\left( t\right)   =\rho^{\text{eq}}_{n,n}+ \sum_{p=1}^{8}\overline{W}_{n,p}^{{\left( 0\right)}}\exp \left[ -R_{p}^{\left( 0 \right) }\left( t-t_{0}\right) \right] \widetilde{\rho }_{p}\left( t_{0}\right)  \text{,}
\end{equation}
where $R_{p}^{\left( 0 \right) } = - C \lambda_{p}^{\left( 0\right)} $ represent the relaxation rates that characterize the dynamics of the zero order density matrix elements.

To end this  mathematical description, it is necessary to highlight some theoretical characteristics. This algebraic mathematical development points out that higher dimensions of linear systems must be accomplished numerically. Similar observation was made for spins systems with $I>2$ on a theoretical approach \cite{kelly1992}. Also, the dimension of the linear system implies in the degree of a polynomial equation used to solve the problem, which is equivalent to find the roots \cite{becker1982}. Therefore, the dimension of the linear system implies on the number of exponential functions used to monitor the dynamics of each coherence order of the density matrix elements. It is evidenced by similar approaches of  previous  non-exponential description of the relaxation procedure in spin systems with nuclear spin $I>1/2$, which is a consensual point of view \cite{mcdowell1995}. The prefactors of each exponential function depend on the spectral density functions, this was predicted studying  radioactive $^{8}$Li nuclei with spin $I=2$ on a solid L--based alloy \cite{korblein1985}.

\end{widetext}

\end{document}